\documentclass{aastex62}
\usepackage{ifpdf}
\usepackage{graphicx}
\usepackage{blindtext}
\usepackage{hyperref}

\def\kms{km~s$^{-1}$}

\def\kms{\mbox{km~s$^{-1}$}}
\def\cmc{cm$^{-3}$}
\def\cmq{cm$^{-2}$}

\def\Msun{\mbox{$M_\odot$}}
\def\Lsun{\mbox{$L_\odot$}}
\def\Rsun{\mbox{$R_\odot$}}
\def\Vlsr{$v_\mathrm{LSR}$}
\def\Vsys{$v_\mathrm{sys}$}

\def\mic{\mbox{$\mu$m}}

\def\wat{H$_2$O}
\def\amm{NH$_3$}

\def\hcop{\mbox{HCO$^+$}}

\def\CO{$^{12}$CO}
\def\co{$^{12}$CO}
\def\tCO{$^{13}$CO}

\def\NtwoH{N$_2$H$^+$}
\def\HCOP{HCO$^+$}
\def\hcop{HCO$^+$}
\def\HtCOp{H$^{13}$CO$^+$}

\def\Lbol{$L_\mathrm{bol}$}

\def\pbeam{beam$^{-1}$}

\def\V0{\mbox{$V_\mathrm{0}$}}

\def\Vtb{\mbox{$v_\mathrm{t,blue}$}}
\def\Vtr{\mbox{$v_\mathrm{t,red}$}}

\def\Vch{\mbox{$v_\mathrm{ch}$}}
\def\Vchb{\mbox{$v_\mathrm{ch,blue}$}}
\def\Vchr{\mbox{$v_\mathrm{ch,red}$}}

\def\Tb{\mbox{$T_\mathrm{B}$}}

\def\Tr21{\mbox{$T_\mathrm{r,21}$}}

\def\Tb{\mbox{$T_\mathrm{b}$}}

\def\Tex{\mbox{$T_\mathrm{ex}$}}
\def\meanTex{\mbox{$\langle T_\mathrm{ex}\rangle$}}
\def\meanTex{\mbox{$\langle T_\mathrm{ex}\rangle$}}

\def\sgm{$\sigma$}

\def\gf{\mbox{GF\,9-2}}
\def\fc{{\it first core}}
\def\fcs{{\it first cores}}

\def\RadpSimSoltn{$\rho (r)\propto r^{-2}$}
\def\RadpInf{$\rho (r)\propto r^{-3/2}$}
\def\Reff{\mbox{$R_\mathrm{eff}$}}
\def\RefValue{250$\pm$80}
\def\ReffValueUnit{\RefValue\,{\rm AU}}

\def\Mcsm{$M_\mathrm{csm}$}
\def\meanMcsmValue{$\sim 8\times 10^{-3}$}

\def\Lbol{$L_\mathrm{bol}$}

\def\lesssim{\mathrel{\hbox{\rlap{\hbox{\lower4pt\hbox{$\sim$}}}\hbox{$<$}}}}
\def\gtrsim{\mathrel{\hbox{\rlap{\hbox{\lower4pt\hbox{$\sim$}}}\hbox{$>$}}}}
\newcommand{\lw}[1]{\smash{\lower2.ex\hbox{#1}}}

\received{}
\revised{}
\accepted{for publication in ApJ on December 12, 2018}

\shorttitle{A 1000-AU scale outflow from a protostar with an age of $\lesssim$4000 yrs}
\shortauthors{Furuya et al.}

\begin{document}

\title{A 1000 AU Scale Molecular Outflow Driven by \\
a Protostar with an age of $\lesssim$4000 Years}

\correspondingauthor{Ray S. FURUYA}

\author{Ray S. FURUYA}
\affil{Institute of Liberal Arts and Sciences, Tokushima University, Minami Jousanjima-Machi 1-1, Tokushima, Tokushima 770-8502, Japan}\email{rsf@tokushima-u.ac.jp}

\author{Yoshimi KITAMURA}
\affil{Institute of Space and Astronautical Science, Japan Aerospace Exploration Agency, Yoshinodai 3-1-1, Chuo-ku, Sagamihara, Kanagawa 252-5210, Japan}
\email{kitamura@isas.jaxa.jp}

\author{Hiroko SHINNAGA}
\affil{Chile Observatory, National Astronomical Observatory of Japan, 
Current address: Department of Physics, Faculty of Science, Kogoshima University, Korimoto 1-21-35, Kagoshima,
Kogoshima 890-0065, Japan}
\email{shinnaga@sci.kagoshima-u.ac.jp}



\begin{abstract}
To shed light on the \textcolor{black}{early phase of a}
low-mass protostar formation process,
we conducted interferometric observations towards
a protostar \gf\ using the CARMA and SMA.
The observations have been carried out
in the \CO\ $J=3-2$ line and
in the continuum emission at the wavelengths 
of 3.3\,mm, 1.1\,mm and 850 \micron\ with a spatial resolution of
$\approx 400$\,AU.
All the continuum images
detected a single point-like source with a beam-deconvolved effective radius
of \ReffValueUnit\ at the center of the \textcolor{black}{previously known} 
1.1 -- 4.5 \Msun\ molecular cloud core.
A compact emission is detected towards the object
at the {\it Spitzer} MIPS and IRAC bands 
\textcolor{black}{as well as the four bands at the {\it WISE}.}
Our spectroscopic imaging of the CO line revealed that 
the continuum source is driving 
a 1000\,AU scale molecular outflow, including a pair of lobes 
where a collimated ``higher" velocity ($\sim 10$ \kms\ with respect to the
velocity of the cloud) red lobe exists 
inside a poorly collimated ``lower" velocity ($\sim 5$ \kms) red lobe.
These lobes are  \textcolor{black}{rather young} (dynamical time scales of $\sim$ 500 -- 
\textcolor{black}{2000} yrs)
and the least powerful 
(momentum rates of 
\textcolor{black}{
$\sim 10^{-8}-10^{-6}$ \Msun\ \kms\ yr$^{-1}$}
) ones so far detected.
A \textcolor{black}{protostellar} mass of $M_\ast \lesssim \,0.06$ \Msun\ was estimated
using an upper limit of the \textcolor{black}{protostellar} age of 
\textcolor{black}{
$\tau_\ast\lesssim (4\pm 1)\times 10^3$ yrs
and an inferred non-spherical steady mass accretion rate of 
$\sim 1\times 10^{-5}$ \Msun\ yr$^{-1}$.}
\textcolor{black}{Together with results from an SED analysis,
we discuss that the outflow system is driven by a protostar whose
surface temperature of \textcolor{black}{$\sim 3,000$\,K,} and that the natal cloud core
is being dispersed by the outflow.}
\end{abstract}




\section{Introduction}
\label{s:intro}

\subsection{Background}
Because of the prominent progress in observational and theoretical 
studies over the past decades, the early phases of 
the formation of an isolated low-mass star are now certainly well understood.
Nevertheless, our knowledge of \textcolor{black}{its} earliest phase is 
obviously limited in the observational studies.
The theoretical studies in the late 1960s and the 1970s indicated that, 
once a cloud core lost support against the collapse due to its self-gravity,
the core collapses isothermally in a runaway fashion, which is followed by 
an accretion process onto a protostellar core.
The first kernel of a mass produced at the center of the collapsing
cloud core is believed to evolve towards a hydrostatic object which 
becomes opaque to the dust continuum emission, and thus the
cooling by the dust emission becomes inefficient.
Such an adiabatic hydrostatic core is referred to as a first hydrostatic core, 
a \fc\ \citep{Larson69}.
The thermal evolution of \fcs\ has been extensively studied by e.g., 
\citet{Masunaga98} who performed 1D radiative hydrodynamic calculations
to track their evolution.
Once the central temperature is elevated to about 2000\,K,
corresponding to the volume mass density of $\rho\sim 10^{-8}$ g \cmc,
molecular hydrogen commences being dissociated into atomic hydrogen.
The dissociation uses the thermal energy which originates from
the released gravitational energy due to the accretion.
After the dissociation is completed, a {\it second hydrostatic core} forms.
The {\it second cores} are believed to correspond to the protostars
observed as ``class 0" objects \citep{Andre93}.
Although feasibility of detecting the \fcs\ have been long discussed 
\citep[e.g.,][]{Boss95, Omukai07, TT11, Commercon12, Tomida13}, 
it is an enormously difficult task for observers to identify the \fcs\ 
simply because they are very short-lived objects.

Despite such difficulty, the growing numbers of the \fc\ candidates
have been reported 
\citep[e.g.,][]{Belloche06, Enoch10, Chen10, Dunham11, Pineda11, Tsitali13, Pezzuto12, 
Hirano14, Friesen14, Maureira17}
based on unbiased surveys which observed the thermal dust continuum emission 
at submillimeter (submm)
to infrared (IR) bands with space telescopes
together with bolometer cameras on ground-based
millimeter (mm) to submm telescopes.
The theoretical studies predicted that the duration of the \fc\ phase should
continue at most a few times $10^3$ yrs 
which is an order of magnitude shorter than the
lifetime of the class 0 objects \citep[e.g.,][]{Andre00}.
Therefore only a handful of the \fc\ objects should exist in
nearby molecular clouds when we consider the number of the 
class 0 sources identified so far and the relative duration between 
the first and second core phases \citep{Masunaga98, Masunaga00a}.
Therefore the number of the candidates is too many 
to be consistent with this statistical argument.
This raises a question that some of the candidates 
may have been misidentified and/or
the duration of the \fcs\ may be longer than the theoretical predictions.
We therefore believe that it is still required to \textcolor{black}{
identify protostars at its early evolutionary stage, and study their physical properties.}

\subsection{Previous Observations}
In this context, we performed a detailed study of 
the natal cloud core harboring 
an extremely young low-mass protostar \gf\ 
at a distance ($d$) of 200\,pc 
\citep[][see a summary in Poidevin \& Bastien 2006]{Wiesemeyer97}
located in the GF\,9 filament \citep{Schneider79}.
The filament was studied by 
near-infrared (NIR) extinction \citep{Ciardi98}, and 
optical and NIR absorption polarization \citep{PB06} observations.
\citet{PB06} showed that there exist
well aligned pc-scale magnetic fields whose overall direction appears
to be almost perpendicular to the filament, 
and discussed that the magnetic fields must have regulated the 
formation and evolution of the filament.
In the filament there are almost equally spaced seven dense cloud cores
traced by the \amm (1,1) lines
\citep[][hereafter Paper II]{rsf08}.
\textcolor{black}{
Subsequently, \citet{rsf14} [hereafter Paper IV] studied the physical properties of the 
tenuous ambient gas surrounding the dense cloud core \gf\
by observing the $J=$1--0 transitions of \CO, \tCO\ and C$^{18}$O molecules,
which covered $\sim$one-fifth of the whole filament.
We found that the filament around the core is supported by turbulent and
magnetic pressures against self-gravity, and argued that
the core has formed through gravitational collapse triggered 
by the local decay of the supporting force(s).}\par

\textcolor{black}{
The cloud core \gf\ has been studied in various molecular lines by many authors.
Using the NIR extinction map together with 
the $^{13}$CO (1--0) and CS (2--1) line data, 
\citet{Ciardi98, Ciardi00} identified 
an $\sim 5$--11\,\Msun\ molecular {\it clump} of the ``GF\,9-Core" 
with a size of $\sim$0.8\,pc$\times$0.6\,pc, 
which harbors the dense cloud core of our interests, 
a {\it protocore} of the ``Southwestern Condensation" \citep{rsf14},
and the IRAS point source PSC\,20503+6006 
\citep[see the left panel of Figure 10 in][]{Ciardi00}.
The dense cloud core is cross-identified as 
L1082\,C \citep[e.g.,][and references therein]{Bontemps96, Caselli02}.
\citet{Caselli02} estimated that the core has 
a mass of 0.35$\pm$0.1 \Msun\ over a radius $\sim 0.14$\,pc 
through the \NtwoH\ (1--0) line observations using the FCRAO 14\,m telescope
($\theta_\mathrm{HPBW} = 54\arcsec$).
Figure 2 in \citet{Caselli02} and Figure 10 in \citet{Ciardi00}
clearly show that the IRAS point source is located at the 
$\sim$90\arcsec\ (corresponding to 0.08\,pc at $d =$\,200\,pc) south-south-west position from the core center.
\citet{Bontemps96} observed the dense cloud core L1082\,C, i.e., \gf\ in 
the CO (2--1) line using 
Caltech Submillimeter Observatory (CSO) 10.4\,m telescope 
($\theta_\mathrm{HPBW} = 30\arcsec$)
as an outflow survey towards low-mass embedded young stellar objects (YSOs).
However, no molecular outflow was detected with 
the upper limit of 1.5$\times 10^{-6}$ \Msun\ \kms\ yr$^{-1}$
in outflow momentum rate.
In contrast to the negative detection of an outflow,
\citet{rsf03} detected the \wat\ maser emission at 22\,GHz towards
the core center using the Nobeyama 45\,m telescope.
Although the beam size of the maser observations was $75\arcsec$
(corresponding to 0.072\,pc at $d =$\,200\,pc), 
the presence of the masers strongly suggests that 
star formation activity has already commenced 
within a radius of 0.036\,pc centered on the cloud core.
Subsequently, we carried out higher resolution observations:
the \NtwoH\ (1--0) and \HtCOp\ (1--0) lines 
using Nobeyama 45\,m telescope ($\theta_\mathrm{HPBW} \simeq 18\arcsec$)
and Caltech Owens Valley Radio Observatory (OVRO) millimeter (mm) array
(synthesized beam size $\theta_\mathrm{syn} \simeq 5\arcsec$)
to study the cloud core \citep[][hereafter Paper I]{rsf06}, 
HCO$^+$ (3--2) line using CSO 10.4\,m telescope (25$\arcsec$)
to detect gas infall in the core \citep[][hereafter Paper III]{rsf09}, and
CO (1--0) and (3--2) lines using Nobeyama 45\,m (17\arcsec) 
and CSO 10.4\,m telescopes (22\arcsec),
respectively, to assess the negative detection of the outflow (Paper I).
Here we scaled the {\it clump} mass and size estimated by \citet{Ciardi98}, 
the cloud core mass  and size by \citet{Caselli02},
and the outflow momentum rate by \citet{Bontemps96} 
using $d =$\,200 pc \citep{Wiesemeyer97}
because \citet{Bontemps96, Ciardi98, Ciardi00, Caselli02} adopted $d=$\,440 pc.
Hereafter, we adopt the distance to the object of $d =$\,200 pc.
}

In Paper I, we argued that the central object(s) deeply embedded 
in the \gf\ core has (have) not generated an extensive molecular outflow, 
\textcolor{black}{as also suggested by \citet{Bontemps96}.
The absence of an extensive outflow} 
should provide us with a rare opportunity to investigate 
the physical properties of the natal core free from the disturbance by the outflow.
\textcolor{black}{In Paper III, we reported that}
``blueskewed asymmetry profiles" in the optically thick \hcop\ (1--0), (3--2),
and HCN (1--0) lines were detected 
\textcolor{black}{within a radius of $r\simeq 30\arcsec$
(corresponding to 0.03\,pc)}
suggestive of the presence of large-scale gas infall.
The estimated infall velocity was shown to have 
reasonable consistency with the predictions 
from the runaway collapse model \textcolor{black}{in the accretion phase}
\citep[][hereafter the LPH solution]{Larson69, Penston69, Hunter77, WS85}, 
and an 
\textcolor{black}{
infall
} 
rate of $\dot{M}_\mathrm{inf} \simeq 2.5\times 10^{-5}$ \Msun\ yr$^{-1}$ 
was deduced (Paper III).
On the other hand, analyzing the \NtwoH\ (1--0) and \HtCOp\ (1--0) line images
that were produced by combining the visibility data taken with the
OVRO mm-array and the 45\,m telescope, 
we revealed that a density profile of \RadpSimSoltn\ 
holds over the annulus of 0.003\,$\lesssim\,r/\mathrm{pc}\lesssim\,$ 0.08.
\textcolor{black}{Considering possible effects by a putative binary system
which was suggested by the positional difference between the 3\,mm
continuum source and the peak position of the \NtwoH\ emission
($\simeq 6\arcsec$; Paper I), we set an inner radius for the region 
showing the \RadpSimSoltn\ profile to be 0.006\,pc.
This inner radius gives} 
an upper limit of the protostar's age of 
$t_\mathrm{protostar}\lesssim 5\times 10^3$ yrs (Paper I).
All the results strongly suggest that the \gf\ core has undergone gravitational
collapse from the initially unstable state, and has just formed a protostar(s)
in the past $\lesssim 5\times 10^3$ yrs in its center.
In order to assess whether or not the protostar has launched a
compact molecular outflow(s) and to search for its driving source(s),
we performed interferometric observations of the 
central region of the core in the \CO\ (3--2) line and continuum emission.\par

\section{Observations and Data Retrieval}
\label{s:obs}
To resolve the issues summarized in \S\ref{s:intro}, we carried out mm- and submm-
wavelength interferometric observations with the
Combined Array for Research in Millimeter-wave Astronomy (CARMA)\footnote{
Support for CARMA construction was derived from the Gordon and Betty Moore Foundation, the Kenneth T. and Eileen L. Norris Foundation, the James S. McDonnell Foundation, the Associates of the California Institute of Technology, the University of Chicago, the states of California, Illinois, and Maryland, and the National Science Foundation. 
CARMA development and operations were supported by the National Science Foundation under a cooperative agreement, and by the CARMA partner universities. } 
(\S\ref{ss:obs_CARMA}) and
the Submillimeter Array (SMA)\footnote{
The Submillimeter Array is a joint project between the 
Smithsonian Astrophysical Observatory and the Academia Sinica Institute of Astronomy and Astrophysics and is funded by the Smithsonian Institution and the Academia Sinica.}(\S\ref{ss:obs_SMA}), respectively.
All the key parameters of the continuum and molecular line emission observations
are summarized in Table \ref{tbl:obs}.
Note that the SMA observations at wavelengths of 840 \micron\ and 1.1\,mm, respectively, provided us with
2.0 and 1.8 times higher angular resolution images than the CARMA did at 3.3\,mm, 
while the CARMA 3.3\,mm observations gave us 3.0 and 2.3 times larger fields of view (FoV)
and could detect 5.4 and 4.0 times larger spatial structures than the SMA 840 \micron\ and 1.1\,mm
observations, respectively. 
\textcolor{black}{
In addition, we used archival {\it Spitzer} Space Telescope\footnote{
This work is partially based on the data taken with {\it Spitzer} Space Telescope, which is operated by the Jet Propulsion Laboratory, California Institute of Technology under a contract with NASA.}
\textcolor{black}{and {\it Wide-field Infrared Survey Explorer}}\footnote{This publication makes use of data products from the Wide-field Infrared Survey Explorer, which is a joint project of the University of California, Los Angeles, and the Jet Propulsion Laboratory/California Institute of Technology, funded by the National Aeronautics and Space Administration.} data.
to verify the presence of
mid-infrared to far-infrared emission towards the protostar
(\S\ref{ss:obs_IR}).
}

\subsection{CARMA Observations} 
\label{ss:obs_CARMA}
The aperture synthesis observations \textcolor{black}{tuned} at 91.18\,GHz,
which is the \textcolor{black}{middle} frequency of between the \NtwoH\ (1--0) and \HCOP\ (1--0) emission,
were carried out using the CARMA with C, D, and E configurations (project code: c0296).
The visibility data used to produce a 3\,mm continuum emission image were obtained 
by 5, 4, and 5 partial tracks performed in 2009 May, April, and June, respectively.
Since our primary goal of the observations is to image these molecular lines by 
combining with the single-dish data taken with the Nobeyama 45\,m telescope (Paper I),
we carried out nineteen-field mosaic observations.
However, in this paper, we concentrate on discussing the continuum emission,
and the line data will be published in another paper.\par

We used J1849+670 as phase and gain calibrators,  and
3C454.3, 3C345, J1642+689, and J1751+096 as passband calibrators.
The flux densities of J1849+670 were determined from observations
of Mars and MWC\,349. The uncertainties of our flux calibration were
estimated to be approximately $10\%$.
The visibility data were calibrated and edited using the MIRIAD package.
For the continuum data, we merged the visibilities in both of the sidebands
to improve the image sensitivity:
the representative frequency of the continuum image was set to be 
the center frequency of the dual bands.
The image construction was done using the MIRIAD package.

\subsection{SMA Observations}
\label{ss:obs_SMA}
The aperture synthesis observations 
were made using the SMA
at the wavelengths ($\lambda$) of 1.1\,mm and 840 \micron\
on 2010 August 10 and 11, respectively
(project code: 2010A-S56).
The observations were in the Compact-North configuration
using the eight and six antennas at  $\lambda\,=$1.1\,mm and 840 \micron, respectively.
During the observations, the optical depth of the terrestrial atmosphere at 
225\,GHz ($\lambda\,=$\,1.3\,mm) measured with the CSO monitor was fairly stable
($\tau_{225}\sim\,0.055$) except for the last three hours in the 
dawn of the first night ($\tau_{225}\sim\,0.070$).
The center frequencies of receivers were tuned at 279.5\,GHz and 356.7\,GHz
to receive not only the continuum emission 
but also molecular lines such as $^{12}$CO (3--2),
\NtwoH\ (3--2) and HCO$^+$ (4--3) at the 840 \micron\ band as well as
HCO$^+$ (3--2) at the 1.1\,mm band.
Note that the \NtwoH\ and HCO$^+$ lines are typical high-density gas tracers.
We configured the correlator so that we can utilize the maximum bandwidth 
of 4 GHz in each polarization for the continuum detection.\par

For phase and amplitude calibrators,  
we observed 3C\,418
and BLLac, whose angular distances to the source are 9.2 and 21 degrees, respectively,
every 5 minutes.
The calibration of the passband response function was done by 
observing 3C\,279, 3C\,454.3, 3C\,345, and 3C\,84 at the beginning and ending of the observations.
The scale factors for converting into the absolute flux densities were
determined by observations towards Neptune and Uranus.
All the visibility data were calibrated and edited using the MIR package.
Image construction was subsequently done with the MIRIAD package.
For the continuum emission data, we produced an image at each sideband
by concatenating the dual polarization data.

\subsection{Spitzer Space Telescope Archive Data} 
\label{ss:obs_IR}
\textcolor{black}{
We retrieved infrared (IR) images towards the 3\,mm continuum source (Paper I)
from the {\it Spitzer Science Center} data archive.
These images are taken at wavelengths of 3.6, 4.5, 5.8, and 8.0 \micron\ taken with 
the {\it Infrared Array Camera} \citep[IRAC; ][]{Fazio04} 
(data ID\#r26296832),
and 24 and 70 \micron\ with the Multiband Imaging Photometer for Spitzer 
\citep[MIPS; ][]{Rieke04}
 (data ID\#r26297088)
on board the {\it Spitzer} Space Telescope.}
\textcolor{black}{
The point spread function (PSF) sizes of the Spitzer IRAC and MIPS
images ranges between} 
\textcolor{black}{1\farcs4 at the 3.5 \micron\ band, 
which are comparable to those in our interferometric observations
and 18\arcsec\ at the 70 \micron\ bands.}
\textcolor{black}{These fully calibrated data were used without spatial smoothing,
and 1$\sigma$ noise levels over emission-free regions are
$2.4\times 10^{-2}$,
$2.6\times 10^{-2}$,
$9.8\times 10^{-2}$,
$0.12\times 10^{-2}$,
$0.13\times 10^{-2}$, and
$0.71\times 10^{-2}$ MJy sr$^{-1}$ for the 
3.6, 4.5, 5.8, 8.0, 24 and 70 \micron\ band data, respectively.
}

\subsection{WISE Archive Data} 
\label{ss:obs_WISE}
\textcolor{black}{
We retrieved infrared (IR) images towards the 3\,mm continuum source 
from the {\it Wide-field Infrared Survey Explorer} \citep[WISE;][]{Wright10} 
data archive.
These images are taken at wavelengths of 3.4, 4.6, 11.6, and 22.1 \micron.
The PSF sizes of the WISE images are $\sim 6\arcsec$ at the 3.4, 4.6 and 11.6 \micron\ bands, 
whereas $\sim 12\arcsec$ at the 22 \micron\ band, which are $\sim$2--4 times larger than
those of the {\it Spitzer} images.
}

\section{Results and Analysis}
\label{s:res}

\subsection{Continuum Emission at 3.3\,mm, 1.1\,mm, and 850 \micron\ Bands}
\label{ss:cont}
In this subsection we describe results \textcolor{black}{obtained} from the mm and
submm continuum 
emission maps (Figure \ref{fig:contmaps}),
which are followed by an analysis of a continuum spectrum (Figure \ref{fig:contsp}).\par

\subsubsection{Maps and Flux Measurements}
\begin{figure*}
\includegraphics[angle=0,scale=0.65, bb=0 0 12cm 8cm]{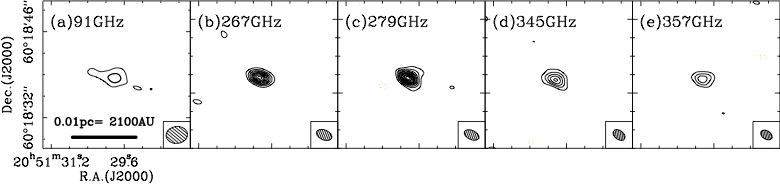}
\caption{
Interferometric continuum emission maps towards the
center of the \gf\ low-mass star forming core at frequencies of 
(a) 91\,GHz ($\lambda\,=\,$3.3\,mm) taken with the CARMA, 
(b) 268\,GHz (1.12\,mm),
(c) 280\,GHz (1.07\,mm),
(d) 345\,GHz (870 \micron), and
(e) 357\,GHz (840 \micron) taken with the SMA.
All the contours, except for the central thick one in (a), 
are drawn with the $3\sigma$ intervals, starting from the $3\sigma$ levels, 
where the $1\sigma$ levels mean the RMS noise levels of the images.
The single thick contour in the panel (a) represents the 5\sgm\ level.
The horizontal bar in (a) indicates the linear size scale of 0.01\,pc
assuming the distance to the source to be 200\,pc.
See Table \ref{tbl:obs} for the image sensitivity and the 
synthesized beam sizes which are indicated by the hatched ellipses at 
the bottom-right corners of the five panels.}
\label{fig:contmaps}
\end{figure*}

The 3.3\,mm map obtained by CARMA \textcolor{black}{(Figure \ref{fig:contmaps}a) was
produced by concatenating the visibility data taken at the two sidebands (\S\ref{s:obs}),
whereas the 1.1\,mm and 850 \micron\ ones by SMA (Figures \ref{fig:contmaps}b - e)
were produced in each sideband.}
At all the bands, 
we detected single point-like sources whose
peak positions agree with each other within the spatial resolutions at 
R. A. $=\,20^h \,51^m \,29.86^s$,
Decl $=60\degr \,18\arcmin \,38\farcs{23}$ in J2000.
In the CARMA 3.3\,mm image the object shows a weak elongated structure to the east.
No other significant emission was detected over the FoVs (Table \ref{tbl:obs}) 
at all the bands.
\textcolor{black}{This would exclude the ``protobinary" hypothesis 
which we discussed in \S5.3 in Paper I.}
\par

We performed photometry of the continuum emission
using the images at the six frequencies by including the 
3.3\,mm data analyzed in Paper I.
\textcolor{black}{
To estimate the flux densities,
we used task \texttt{JMFIT} in AIPS package and \texttt{IMFIT} in CASA 
to perform beam-deconvolution.}
Table \ref{tbl:phot} summarizes our flux density ($S_\nu$) measurement at each frequency
with an assumption that 
the morphology of the object can be approximated by a 2D elliptical Gaussian.
A comparison of the synthesized beam sizes (Table \ref{tbl:obs})
and the beam-deconvolved source sizes
\textcolor{black}{(see the $\Theta_\mathrm{maj}\times\Theta_\mathrm{min}$ values 
in Table \ref{tbl:phot})} indicates
that our observations \textcolor{black}{barely resolved the emission at all the bands,}
\textcolor{black}{i.e., they are detected as slightly extended sources.}\par

After assessing the beam-deconvolved sizes, 
we measured the area of the emanating region in the plane-of-sky
($A_\mathrm{s}$) to calculate its effective radius (\Reff) by
$A_\mathrm{s}\,=\,
\frac{\pi}{4\ln 2}(\Theta_\mathrm{maj}\times\Theta_\mathrm{min})
\,\equiv\,\pi R_\mathrm{eff}^2$.
For further analysis, we use the mean \Reff\ of
\ReffValueUnit\ calculated from the SMA bands of 1.1\,mm and 850 \micron\ 
\textcolor{black}{because the synthesized beam size ($\theta_\mathrm{syn}$) 
of $\sim 2\arcsec$ at SMA
observations is better than those at 3\,mm.}\par

\textcolor{black}{In order to compare the CARMA 3\,mm measurement
with the previous one at 
3\,mm using the OVRO mm-array 
($\theta_\mathrm{syn}\sim 5\arcsec$; Paper I),}
\textcolor{black}{we checked consistency between the photometric results from 
the above Gaussian-fitting method and those measured over 
the region enclosed by the 3\sgm\ level contours using the CARMA and SMA data.
This is because we adopted the latter method in Paper I.
We found a reasonable consistency between the two methods, 
suggesting} that the 2D elliptical Gaussian 
\textcolor{black}{approximation adopted in the beam-deconvolution
processes is reasonable.}\par

\subsubsection{Uncertainty in the Flux Measurements}
\label{ss:mmFluxUncertainty}
\textcolor{black}{
Based on the minimum spatial frequencies of our observations (Table \ref{tbl:phot})
and the discussion in Appendix of  \citet[][]{WW94},} 
we estimated that our \textcolor{black}{SMA} observations 
missed approximately 70\% of the flux densities for
the extended components of the envelope
with respect to the expected zero-spacing flux densities
assuming a model with a power-law radial density profile (Paper I).\par

\textcolor{black}{Table \ref{tbl:phot} presents the mm- and submm continuum flux densities where we included the previous OVRO measurement (Paper I).
We verified that the 3\,mm flux difference 
between the OVRO and CARMA measurement is real. 
We qualitatively argue that the difference would be caused in the process
of synthesis imaging due to the various differences, e.g., 
those in spatial-frequency ($D_\lambda$) ranges 
(3.8 $\leq D_\lambda/k\lambda \leq$\,67 for OVRO vs. 1.93 $\leq D_\lambda/k\lambda \leq$\,101.7 for CARMA) and samplings in the $(u, v)$ coverage, 
their weighting functions (the OVRO data were imaged with natural weighting,
while the CARMA data with ``Robustness of +2"),
correlator bandwidths (4096\,MHz for OVRO vs. 938\,MHz for CARMA),
usage of multi-frequency synthesis method in CARMA,
and the beam sizes adopted for the photometry (5\farcs38$\times$4\farcs80 for OVRO
vs. 3\farcs89$\times$3\farcs84 for CARMA).
Among these possible causes, we argue that the difference in the $D_\lambda$ ranges
may be the most dominant one. However, it is not trivial to explain quantitatively
the photometric difference between 
the OVRO and CARMA results.
}

\subsubsection{Continuum Spectrum over the Millimeter and Submillimeter Bands}
\label{sss:ContSpIndex}
\begin{figure}
\includegraphics[angle=0,scale=.24, bb=0 0 15cm 10cm]{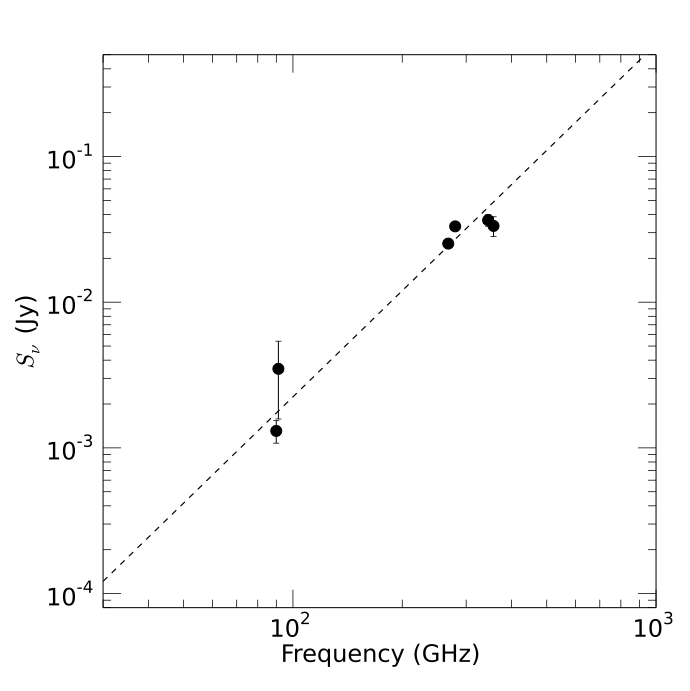}
\caption{
Plot of the \textcolor{black}{mm and submm} 
continuum spectrum of the compact source embedded in the \gf\ cloud core.
The flux densities presented in this panel are summarized in Table \ref{tbl:phot}.
The broken-line indicates the best-fit power law spectrum
with a form of $S_\nu \propto \nu^\alpha$
where the best-fit spectral index is $\alpha = 2.4 \pm 0.3$ (\S\ref{ss:cont}).
\label{fig:contsp}}
\end{figure}

Figure \ref{fig:contsp} presents an interferometric mm- and submm continuum spectrum 
including the OVRO measurements.
\textcolor{black}{Compared to the 1.1\,mm and 850 \micron\ flux densities, 
those at  3.3\,mm drop almost one order of magnitude 
(Table \ref{tbl:phot}; see also Figure \ref{fig:contsp}).}
Assuming the power-law spectrum with
$S_\nu = S_0 \,\nu^\alpha$, we obtained 
the best-fit spectral index of $\alpha = 2.4\pm 0.3$.
\textcolor{black}{If we assume that the observed continuum emission at
mm and submm regime} is fully attributed to 
optically-thin thermal dust emission represented by a single-temperature 
graybody emission and that
the dust mass absorption coefficient can be written by $\kappa_\nu\propto \nu^\beta$,
the best-fit $\alpha$ value 
\textcolor{black}{obtained between 
the wavelength range of $\lambda\,=$\,3\,mm -- 850\,\micron}
leads to the $\beta$-index of $0.4\pm 0.3$.
The inferred $\beta$ is comparable to those measured
at the wavelength range
towards circumstellar disks around class II sources 
($\beta\lesssim$ 1.0) rather than those measured in 
molecular clouds,
preprotostellar cores, and
circumstellar envelope associated with protostars ($\beta\sim$ 1.5 -- 2.0)
\citep[e.g.,][]{Beckwith00, AW07a, AW07b, Ricci10a, Ricci10b, PC11b}.
\textcolor{black}{Clearly the estimated $\beta$-index is smaller than those
expected for an envelope harboring a protostar.
This yields
that there might be forming an extremely-dense compact optically-thick region
around the protostar or there could still exist a remnant of a first hydrostatic core
\citep[e.g.,][]{Larson69}(described in \S\ref{sss:Geometry_by_Ages}).}

It is theoretically shown that small dust grains can grow
in the innermost densest part of infalling envelopes, 
e.g., \citet{Hirashita09, Ormel11}.
We therefore estimate mass of the \gf\ envelope to infer a mean volume density ($n$)
to see such a possibility.
\textcolor{black}{Assuming that the observed continuum emission is 
attributed to optically thin thermal dust emission,}
the total mass of the circumstellar materials, \Mcsm, can be \textcolor{black}{calculated} by 
\Mcsm $=\frac{S_\nu d^2}{\kappa_\nu B_\nu(T_\mathrm{d})}$ where
$\kappa_\nu$ is the dust mass absorption coefficient at frequency $\nu$, and 
$B_\nu(T_\mathrm{d})$ the Plank function with dust temperature of $T_\mathrm{d}$.
In these calculations, we adopted the usual assumption that 
$\kappa_\nu$ has a form of $\kappa_0(\nu/\nu_0)^\beta$ where
$\kappa_0$ is a reference value at a reference frequency, $\nu_0$. 
We used the $\kappa_0 =$ 0.1 cm$^2$ g$^{-1}$ 
at $\nu_0\,=$\,1.2\,THz with $\beta\,=$\,1.8 \citep{PC11b}.
Notice that the $\kappa_0$ and $\beta$ values lead to
$\kappa_\mathrm{231~GHz}$ of 0.007 cm$^2$ g$^{-1}$ 
which falls between the two values used in Paper I.
Moreover we assumed that the dust in the region is well-coupled with the gas.
Hence we considered that the excitation temperature (\Tex) of the 
\NtwoH\ (1--0) lines (Paper I) represents the gas ($T_\mathrm{gas}$)
and dust ($T_\mathrm{d}$) temperatures, i.e., 
\Tex\ $=\,T_\mathrm{gas}\,=\,T_\mathrm{d}$.
We adopted the mean excitation temperature of \meanTex\ $=\,22.6\pm 3.6$\,K 
which is the mean value measured inside the 3$\sigma$ level contour of the 840\,\micron\
continuum emission in Figure \ref{fig:contmaps}e (Appendix \ref{as:Tex}).
We calculated \Mcsm\ values for the individual bands (Table \ref{tbl:phot});
the mean value of \meanMcsmValue\ \Msun\ for the higher resolution SMA results
and the \Reff $=$\ReffValueUnit\ leads to the order of
$n$ is $10^7$ \cmq.
This may be too low for dust grains to coagulate, hence 
it is unlikely that dust grains over the entire envelope has already grow
as in more evolved objects.\par

\subsection{\textcolor{black}{Continuum Emission in the Infrared Image Data}}
\label{ss:IRcont_results}

\subsubsection{\textcolor{black}{Spitzer  data}}
\begin{figure}
\includegraphics[angle=0,scale=.1, bb=0 0 15cm 10cm]{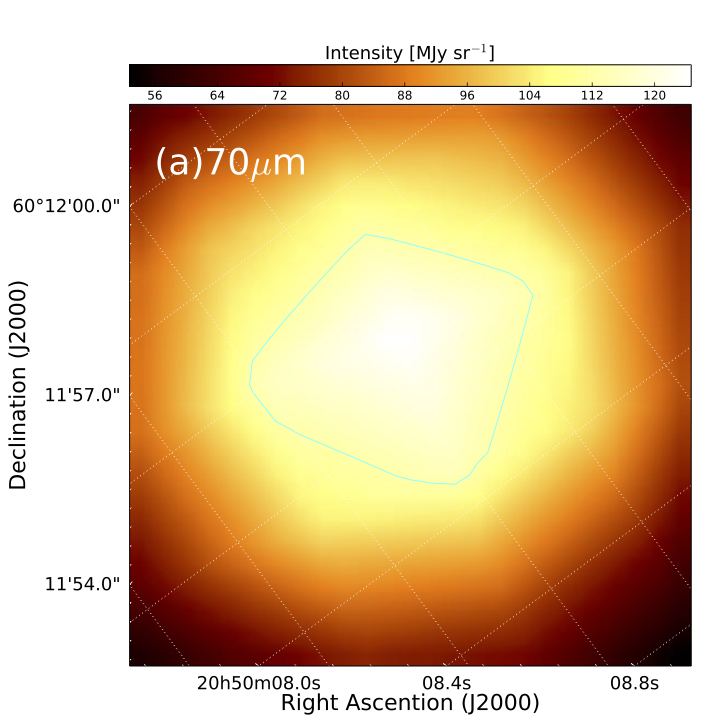} 
\caption{\textcolor{black}{
Continuum emission images in unit of MJy sr$^{-1}$ towards the
center of the \gf\ core at wavelength of 
(a) 70\,$\mu$m, 
(b) 24\,$\mu$m, 
(c) 8.0\,$\mu$m, 
(d) 5.8\,$\mu$m, 
(e) 4.5\,$\mu$m, and
(f) 3.6\,$\mu$m
taken with the {\it Spitzer} 
MIPS filters [(a) and (b)] and
IRAC filters [(c)--(f)].
The center positions of all the images are 
\textcolor{black}{reprojected to}
the position of the 840\,$\mu$m (357\,GHz) source (\S\ref{ss:cont}).
The cyan contour in each panel presents the 90\% level one with respect to the peak intensity 
in each image.
The RMS noise levels of the {\it Spitzer} images
are described in \S\ref{ss:IRcont_results}.}
\label{fig:SpitzerMaps}}

\end{figure}

In Figure \ref{fig:SpitzerMaps}, 
we present the {\it Spitzer} IRAC and MIPS images centered on the 
peak position of the SMA 357\,GHz source
(Figure \ref{fig:contmaps}e).
\textcolor{black}{Assuming that 
the absolute positions in the IRAC and MIPS images
agree with each other within 0\farcs36,
which is the pixel size common to all the $Spitzer$ images,
these images were ``registered" to the 3.6 \mic\ one by ``reprojecting" using
\texttt{astropy} package.
Subsequently these images were compared to the SMA images, 
whose positional accuracies are estimated to be
approximately $\sim 10\%$ of the synthesized beam sizes of 
0\farcs2 based on the typical errors of the interferometric baseline-vector measurements, 
angular separations to the calibrators, and their absolute positions.}
Clearly a point source is detected towards the position of the submm source
at all the IR bands.
We verified that no other infrared sources are detected
within a radius of $\sim 100\arcsec$ (corresponding to $\sim$0.1\,pc)
at the 70, 24, and 8 \mic\ bands with the detection threshold of the 3\sgm\ level in each image.\par

The flux density at each band was computed from the image in unit of
MJy sr$^{-1}$ with aperture
photometry using task \texttt{apphot} in IRAF package with a standard manner;
we integrated the emission inside an optimized circle centered on the source
after subtracting the background.
Notice that the 70 \mic\ flux density is considered as upper limits
because the HPBWs of the aperture encompassing the source is
significantly larger than that of interest for us.
The photometry results are summarized in Table \ref{tbl:apphot}.

\subsubsection{\textcolor{black}{WISE data}}
\textcolor{black}{
Towards the 3\,mm continuum source, we clearly detected a point-like source
in all the four {\it WISE} bands.
The object is identified as \texttt{WISE~J205129.83+601838} in the 
WISE All-Sky Release Source Catalog
from which we obtained its Vega magnitudes.
These magnitudes were converted into flux densities in unit of Jy (Table \ref{tbl:apphot})
with an equation of $F_\nu = F_{\nu,0}\times 10^{(-m_\mathrm{Vega}/2.5)}$
where $F_{\nu,0}$ is a zero magnitude flux density \citep[][]{Wright10} and $m_\mathrm{Vega}$
calibrated WISE Vega magnitudes of the source.
The resultant fluxes are shown in Table \ref{tbl:apphot}.
}

\subsection{$^{12}$CO (3--2) Line Emission}
\label{ss:lines}
Next we present the results from analysis of the SMA \co\ (3--2) line data
\textcolor{black}{(Figure \ref{fig:chmaps})} and
\textcolor{black}{from re-analysis of the previously published \co\ (3--2) line data taken with
the Caltech Submillimeter Observatory (CSO)\footnote{The Caltech Submillimeter Observatory was operated by the California Institute of Technology under the grant from the US National Science Foundation (AST 05-40882). }\,10.4\,m telescope (Paper I).}
We also present an overall picture of the data from
total integrated intensity map \textcolor{black}{(Figure \ref{fig:totmap})} and 
interferometric spectrum \textcolor{black}{(Figure \ref{fig:sp})}
where we detected high-velocity wing emission.
\textcolor{black}{This 
forced us to return to checking the single-dish spectra 
\textcolor{black}{(Figures \ref{fig:CSOvsSMA} and \ref{fig:pmap})}.
After these data inspections, we argue that the CO emission is attributed to 
a 1000 AU-scale molecular outflow \textcolor{black}{(Figure \ref{fig:lobemaps})}
driven by the continuum source
(\S\ref{ss:cont} and \S\ref{ss:IRcont_results}).}


\subsubsection{Velocity Channel Maps}
\label{sss:velstructure}
Figure \ref{fig:chmaps} presents 
velocity-channel maps of the \co\ (3--2) emission taken with the SMA. 
The blueshifted emission 
between \Vlsr\ $=\,-5.6$ and $-3.6$ \kms\
is mainly seen towards the southwest of the continuum source, 
although the emission is much weaker than the redshifted one.
Similarly the redshifted emission is also seen 
towards the southwest, and is detected over a wider velocity range
between \Vlsr\ $=\,-2.0$ and $6.8$ \kms.
In the velocity channels where the intense emission
is detected, e.g., the panels between \Vlsr\ $=\,-2.0$ and $+0.8$ \kms, 
the negative contours are seen along the northeast--southwest direction.
We argue that such an artifact is caused by the limited ($u,v$) coverage 
in our SMA observations.\par

\subsubsection{Spectral Line Profile}
\label{sss:spectra}
\textcolor{black}{Averaging the CO emission along the velocity channel 
(Figure \ref{fig:chmaps})
over the region enclosed by the 3$\sigma$ level contour of
the total integrated intensity map (Figure \ref{fig:totmap}), 
we produced an interferometric \CO\ (3--2) spectrum
(Figure \ref{fig:sp}).}
\textcolor{black}{Here we selected only the bright emission 
seen in the central peak of Figure \ref{fig:totmap} to extract the CO emission 
towards the continuum source.} 
The specific intensity scale in unit of Jy \pbeam\ of the spectrum 
was converted to 
brightness temperature (\Tb) in K using the relation of 
\Tb $=\frac{c^2}{2k_\mathrm{B}\nu^2}\frac{1}{\Omega_\mathrm{s}}\int_{\Omega_\mathrm{s}} I_\nu\mathrm{d}\Omega$
where $k_\mathrm{B}$ is the Boltzmann constant,
$c$ the light velocity,
and $\Omega_\mathrm{s}$ the source solid angle.
\textcolor{black}{The \CO\ (3--2) spectrum taken with SMA shows a noticeable line profile.
It has well-defined high velocity wing emission, especially on the redshifted side.}
\par

\textcolor{black}{
\textcolor{black}{Figure \ref{fig:CSOvsSMA}} compares the single-dish and interferometric
\CO\ (3--2) spectra in the \Tb\ scale taken with the CSO 10.4\,m
telescope (Paper I) and the SMA, respectively.
The SMA spectrum was produced by averaging the CO emission 
inside the CSO beam of 
$\Omega_\mathrm{b} = \frac{\pi}{4\ln 2}\theta_\mathrm{HPBW}^2$
($\theta_\mathrm{HPBW} = 22\arcsec$).
On the other hand, the previously published CSO data 
were reprocessed by revising the velocity ranges for the emission-free channels 
to determine the baseline of each spectrum (Figure \ref{fig:pmap}).
Our re-analysis is motivated by the fact that 
we did not know the presence of the high-velocity wing emission (Figure \ref{fig:sp})
when we had reduced the data for Paper I.
This re-analysis allowed us to detect the redshifted tail emission up to 
\Vlsr\ $=$\,3.6 \kms\ and at a single channel centered on \Vlsr\ $=$\,9.2 \kms.}
\textcolor{black}{
Notice that the high-velocity wing emission is detected only towards
the center position (Figure \ref{fig:pmap}).}\par
\textcolor{black}{
Using the spectra shown in Figure \ref{fig:CSOvsSMA},
we computed the integrated intensities of the redshifted emission to be
$\int T_\mathrm{b}\mathrm{d}v = 1.6\pm 0.4$ K\,\kms\ for the CSO
spectrum and $0.61\pm 0.09$ K\,\kms\ for the SMA one 
over a velocity range of $0.2 \leq$\Vlsr /\kms $\leq 7.0$ (described in \S\ref{sss:velstructure}).}
\textcolor{black}{
This difference in the total intensity is likely to be caused by the typical uncertainties 
($\sim 20\%$) in the absolute flux calibrations of the CSO and/or SMA observations, 
and not by the resolve-out effect in the SMA observations (see the LAS values in Table \ref{tbl:obs}). 
In other words, the above integrated intensities have not only the thermal noise quoted above
but also systematic errors.
Contrary to very extended circumstellar envelopes (\S\ref{ss:cont}), 
a molecular outflow driven by a protostar in its early stage 
is generally enough compact to be fully imaged even by the interferometers
having small numbers of element antennas   
\citep[e.g.,][]{Arce06}. 
We again stress that our CSO observations 
with the $\theta_\mathrm{HPBW} = 22\arcsec$ beam
detected the redshifted wing emission only toward the center position (Figure \ref{fig:pmap}).}
\par

The most straightforward interpretation 
\textcolor{black}{
for the high-velocity wing emission}
is that the protostar of interest is driving a molecular outflow.
This is simply because the velocity difference between the systemic
velocity (\Vsys) and the terminal velocity of the redshifted emission
(\Vtr) of
$|\Vtr - v_\mathrm{sys}|\,=$ 9.5 \kms\ is 
\textcolor{black}{
comparable with typical velocities of the outflow lobes
observed in low-mass class 0 sources \citep[an order of 1 -- 10 \kms; e.g., ][]{Bachiller96, Tobin16}, 
and is larger than a typical velocity of the envelope gas motions
identified as rotation 
and/or infall \citep[an order of 0.1 -- 1 \kms; e.g., ][]{Walker86, Mardones97, Momose98}.
However,} the velocity difference on the blueshifted side is moderate:
$|\Vtb - v_\mathrm{sys}|\,=$ 3.3 \kms.
\textcolor{black}{In this paper, we consider that the blue one is
attributed to an outflow system rather than rotation.
}\par

Another clear feature seen in the SMA \CO\ (3--2) spectrum is that 
no-emission is seen around the \Vsys. 
We consider both or either of the following two causes.
One is that the bulk emission near \Vsys\ is extended,
thus it was resolved out by the interferometer observations.
Here the largest detectable angular size scale in our SMA observations
was estimated to be 27\arcsec\ (Table \ref{tbl:obs}),
corresponding to 0.026\,pc at $d\,=$\,200 pc.
\textcolor{black}{
Therefore, the extended ($\gtrsim 0.1$\,pc) low-velocity 
filament gas \citep[][see also Figures 8a and b in Paper I]{rsf14} 
was spatially filtered out by the aperture synthesis observations.}
The other cause is that the bulk emission around \Vsys\ is
optically thick, producing a self-absorption trough in the spectrum.

\subsubsection{Spatial and Velocity Structures of the Outflow Gas}
\label{sss:velstructure}
\begin{figure}

\centering
\includegraphics[angle=-90,scale=.76, bb=0 0 18cm 26cm]{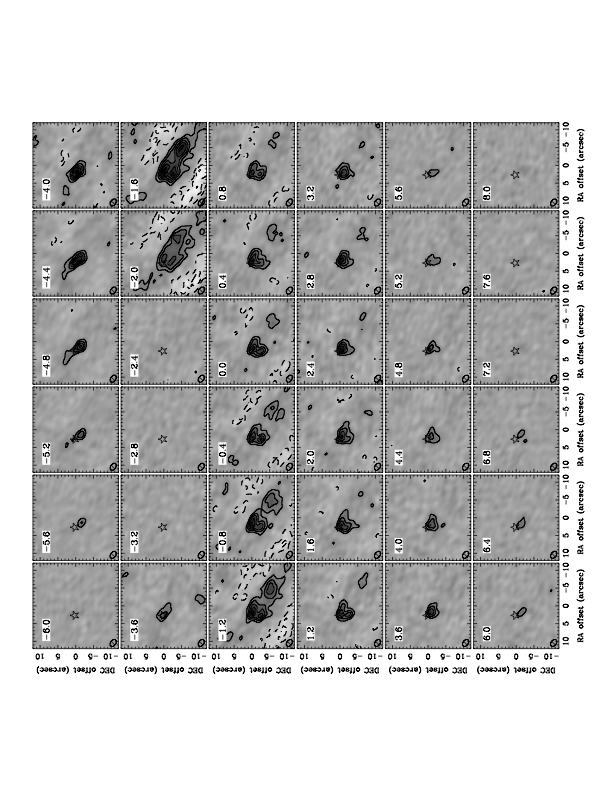}
\vspace{2truecm}
\caption{
Velocity channel maps of the \co\ (3--2) emission 
towards the submm continuum source (Figure \ref{fig:contmaps})
in the \gf\ cloud core.
The stars in these panels indicate the position of the continuum source.
Each channel map is averaged over a 0.4 \kms\ bin whose
central velocity in unit of \kms\ is shown at the top-left corner.
The size of the each panel is 24\arcsec$\times$\,24\arcsec,
corresponding to 4800 AU$\times$4800 AU at $d\,=\,200$.
All the contours are the $3\sigma$ intervals starting from the $3\sigma$
level where $1\sigma$ corresponds to 122 mJy \pbeam.
The systemic velocity of the cloud core is \Vlsr $=-2.48$ \kms\ (Paper I).
\label{fig:chmaps}}
\end{figure}

\begin{figure}
\includegraphics[angle=0,scale=.6, bb=0 0 15cm 15cm]{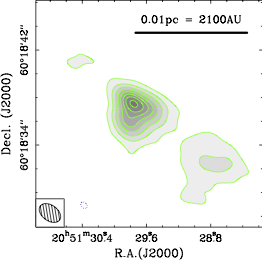}
\caption{
\textcolor{black}{
Total integrated intensity map of the \co\ (3--2) emission
produced from the velocity channel maps (Figure \ref{fig:chmaps}).
The emission is integrated over 
$-5.8\leq v_\mathrm{LSR}/\kms \leq +7.0$ (see Figure \ref{fig:chmaps}).
The green contours with greyscale are plotted with 
the 3$\sigma$ intervals starting from the $+3\sigma$ level where 
the 1$\sigma$ RMS noise level is 
0.62 Jy \pbeam\ \kms.
The synthesized beam size is indicated by the hatched ellipse at 
the bottom-left corner.
The horizontal bar at the top-right corner indicates 
the linear size scale of 0.01\,pc.}
\textcolor{black}{
This map is used for producing
the SMA \co\ spectrum shown in Figure \ref{fig:sp}
}
\label{fig:totmap}}
\begin{center}

\end{center}
\end{figure}

\begin{figure}
\includegraphics[angle=0,scale=.42, bb=0 0 18cm 14cm]{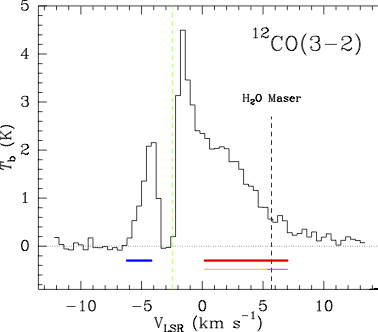}
\caption{
\co\ (3--2) emission spectrum 
\textcolor{black}{towards the center of the \gf\ cloud core observed with the SMA
in brightness temperature (\Tb) scale in units of K
(see \S\ref{ss:lines}).
We made the spectrum by spatially integrating the specific intensity 
of the CO emission in each velocity channel inside a common region enclosed
by the 3\sgm\ contour of a total integrated intensity map (Figure \ref{fig:totmap})
in order to fully detect high-velocity tails.
The solid angle of the 3\sgm\ region is $\Omega_\mathrm{s} = 6.42\times 10^{-10}$ sr,
i.e., an area $A$ of 27.3 arcsec$^2$, 
corresponding to an effective radius of $R_\mathrm{eff} = \sqrt{A/\pi} = 2\farcs9$, i.e., 
580 AU at $d$ of 200\,pc.}
The green and black vertical dashed-lines at \Vlsr $=-2.48$ \kms\ and
$+5.6$ \kms, respectively, show the 
systemic velocity of the cloud (Paper I) and the velocity of the \wat\ maser emission at 22\,GHz \citep{rsf03}.
\textcolor{black}{The horizontal blue and red thick bars under the spectrum indicate
the velocity ranges for producing the outflow lobe maps shown in 
Figure \ref{fig:lobemaps}, and the orange and magenta thin bars under the red one indicate 
those for Figure \ref{fig:redlobemaps}}.
\label{fig:sp}}
\end{figure}

\begin{figure}
\includegraphics[angle=0,scale=.4, bb=0 0 18cm 17.5cm]{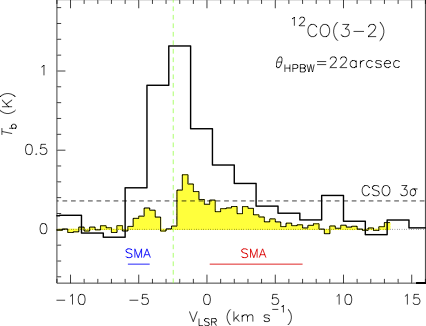}
\caption{Comparison of the \co\ (3--2) line spectra taken with 
the single-dish CSO 10.4\,m telescope (thick histogram) and with the
SMA interferometer (yellow-shaded thin histogram).
Both the spectra are shown in brightness temperature scale in units of K
by averaging over \textcolor{black}{the CSO beam area 
with $\theta_\mathrm{HPBW} = 22\arcsec$}
centered on 
the submm continuum source.
The CSO spectrum towards the center (see Figure \ref{fig:pmap})
was Hanning-smoothed to increase signal-to-noise (S/N) ratio.
The horizontal dashed line indicates the the 3\sgm\ noise level for the CSO spectrum.
The RMS 1\sgm\ noise levels are 60\,mK with a velocity resolution of 1.6 \kms\ 
for the CSO spectrum and
16\,mK with a resolution of 0.4 \kms\ for the SMA one.
\textcolor{black}{The vertical green dashed line indicates the systemic velocity of the cloud (Paper I).
The horizontal blue and red bars under the spectrum are the same as
the thick ones in Figure \ref{fig:sp}, and
indicate the velocity ranges for producing the outflow lobe maps shown in Figure \ref{fig:lobemaps}.
}
Notice that the telescope pointing accuracy in the CSO observations 
was better than 5\arcsec\ (Paper I),
and that the absolute flux calibrations of both the CSO and SMA observations
have uncertainties of $\sim 20\%$.
\label{fig:CSOvsSMA}}
\end{figure}

\begin{figure}
\includegraphics[angle=0,scale=.42, bb=0 0 18cm 20.8cm]{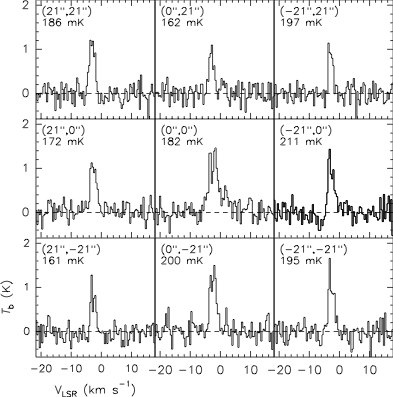}
\caption{Nine spectra of the \CO\ (3--2) line in the main-beam brightness
temperature scale in units of K
observed with a 21\arcsec\ grid centered at the mm-continuum
source of the \gf\ cloud core center. 
The data were previously taken with the CSO 10.4\,m telescope 
(Paper I; \textcolor{black}{$\theta_\mathrm{HPBW} = 22\arcsec$}), 
and are re-analyzed in this study (\S\ref{sss:spectra}).
The telescope pointing position for each spectrum is shown 
at the top-left corner of each panel with an angular-offset form of
($\Delta\alpha, \Delta\delta$) in units of arcsecond.
The 1$\sigma$ RMS noise level of each spectrum in a velocity 
resolution of 0.4 \kms\ is shown in units of mK
below the parenthesis.
\label{fig:pmap}}
\end{figure}

\begin{figure}
\includegraphics[angle=0,scale=.55, bb=0 0 18cm 14.2cm]{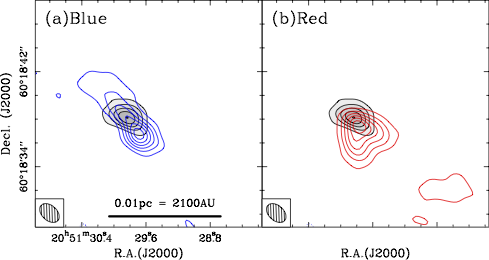}
\caption{Integrated intensity maps of the outflow components 
observed by the \co\ (3--2) emission 
(blue and red contours)
overlaid on the 357\,GHz continuum emission image
(black contours with grayscale map)
(Figure \ref{fig:contmaps}d).
The map size is $21\arcsec\times 21\arcsec$.
The color contours in the two panels show
(a) the blue lobe of $-5.8\leq v_\mathrm{LSR}/\kms \leq -4.2$, and
(b) the red lobe of $+0.2\leq v_\mathrm{LSR}/\kms \leq +7.0$.
\textcolor{black}{See also the blue and red thick bars under the spectrum 
in Figure \ref{fig:sp} for the velocity ranges.}
\textcolor{black}{
All the contours are plotted with the 3$\sigma$ intervals starting from 
the $3\sigma$ levels.
The RMS noise levels of the blue and red maps are 
10.4 and 60 mJy \pbeam\ \kms, respectively.
The ellipse in the box at the bottom left corner of each panel shows
the size of the synthesized beam of the \co\ (3--2) observations.
The linear size scale of 0.01 pc is shown by the bar in the panel (a).
}
\label{fig:lobemaps}}

\end{figure}

Next we examine the spatial and velocity structures of the outflow gas 
unveiled by our SMA \co\
(3--2) line spectroscopic imaging. 
\textcolor{black}{Figure \ref{fig:lobemaps}} presents integrated intensity maps of the compact outflow
lobes overlaid on the 357\,GHz continuum image which 
\textcolor{black}{has the highest angular resolution among our interferometric
continuum images}.
In order to produce the lobe maps, we defined two LSR-velocity ranges
by considering all the features seen in 
the channel maps (Figure \ref{fig:chmaps}) and
the spectrum (Figure \ref{fig:sp}):
the blueshifted emission by
$-5.8\le v_\mathrm{LSR}/\mathrm{km\,s^{-1}}\leq -4.2$ 
\textcolor{black}{
and the redshifted emission by
$+0.2\leq v_\mathrm{LSR}/\mathrm{km\,s^{-1}}\leq +7.0$.
}
After defining the LSR-velocity ranges, we calculated
a characteristic velocity of each lobe ($v_\mathrm{ch}$), which is
an intensity weighted mean LSR-velocity, in order to compare
with other first core candidates in the literature.
\textcolor{black}{This is because
the majority of the previous studies which 
address outflow properties driven by 
first core candidates and very low luminosity objects}
preferred to use $v_\mathrm{ch}$ to represent the flow velocity rather than
using the terminal velocity.
We therefore 
calculated \Vchb $= -4.6$ \kms\ in the LSR-velocity for the blue lobe,
and \Vchr $= +2.5$ \kms\ for the red lobe.\par

\textcolor{black}{
The redshifted lobe is seen to the southwest of the continuum emission,
whereas the blueshifted one spreads towards both southwest and northeast.
This spatial-velocity structure seen in Figure \ref{fig:lobemaps}, 
especially the redshifted one, can be recognized in the early \co\ (3--2) single-dish map 
taken with the CSO 10.4\,m telescope
(see Figure 8d in Paper I).
However, a caution must be used to compare 
Figure 8d in Paper I and Figure \ref{fig:lobemaps}
because the former was obtained for a velocity range of
$-1.4\leq v_\mathrm{LSR}/\mathrm{km\,s^{-1}}\leq +0.6$, 
whereas the latter 
$+0.2\leq v_\mathrm{LSR}/\mathrm{km\,s^{-1}}\leq +7.0$.
Namely, because of the small overlap between the two velocity ranges}
\textcolor{black}{defined in the CSO and SMA data,}
\textcolor{black}{we argue that the Figure 8d in Paper I represents mostly the ambient gas,
hence we do not consider Figure 8d in Paper I for the discussion about
the outflows.
}

\textcolor{black}{ 
Comparing the spatial extent of the blue and red lobes
with that of the 357\,GHz continuum emission (see Figure \ref{fig:lobemaps}), 
we suggest that the ``circum"stellar envelope is, highly likely, 
being evacuated by the outflow.
Such dispersal of the parental gas by outflows has been observed 
at the core-scale}
\textcolor{black}{at 0.1 to 0.01\,pc} 
\textcolor{black}{with single-dish telescopes \citep[e.g., ][]{Monin96, Yoshida10} and
at the envelope-scale with interferometers \citep[e.g., ][]{Momose96, Velusamy98, Arce06}.
This is discussed in \S\ref{ss:StellarMass}.
}


\textcolor{black}{
In the redshifted lobe map, there exist two 
components (Figure \ref{fig:lobemaps}b).
One is associated with the continuum source
(see also the velocity-channel panels of }
\textcolor{black}{
$0.4\lesssim v_\mathrm{LSR}/\kms \lesssim 6.8$
}
\textcolor{black}{
in Figure \ref{fig:chmaps}).
The other is $\sim 8\arcsec$ (corresponding to $\sim 1600$ AU) 
southwest of the continuum source,
although the latter becomes weak in the higher velocity panels
($+1.6\lesssim v_\mathrm{LSR}/\kms \lesssim +3.2$) of Figure \ref{fig:chmaps}.
It is also interesting that the former component shows
a fan-shaped  structure opened to the southwest 
(Figure \ref{fig:lobemaps}b), which is well-recognized
especially in the velocity channel panels of $0.4\lesssim v_\mathrm{LSR}/\kms \lesssim +2.8$
of Figure \ref{fig:chmaps}.
Different from the lower velocity fan-shaped redshifted emission,
the higher velocity redshifted emission in the velocity range of
$+5.6\lesssim v_\mathrm{LSR}/\kms \lesssim +6.8$
is elongated toward the southwest (Figure \ref{fig:chmaps})}.\par

\begin{figure}
\includegraphics[angle=0,scale=.55, bb=0 0 18cm 14.8cm]{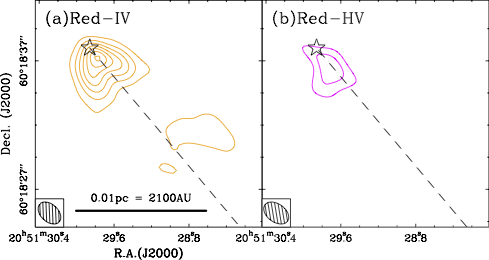}
\caption{
Integrated intensity maps of 
(a) the redshifted intermediate-velocity (``red IV") outflow lobe and 
(b) the redshifted high-velocity (``red HV") one
identified in the redshifted outflow lobe (Figure \ref{fig:lobemaps}b).
The map size is $18\arcsec\times 18\arcsec$.
The ``red IV" lobe map is obtained by integrating the emission over
$+0.2\leq v_\mathrm{LSR}/\kms \leq +5.4$ and 
the ``red HV" one over $+5.4< V_\mathrm{LSR}/\kms \leq +7.0$.
\textcolor{black}{See also the orange and magenta thin bars under the spectrum 
in Figure \ref{fig:sp} for the velocity ranges.}
All the contours are plotted with the 3$\sigma$ intervals starting from 
the $+3\sigma$ level.
The RMS noise levels of the ``red IV" and ``red HV" maps are 
334 and 60 mJy \pbeam\ \kms, respectively.
The star marks the peak position of the submm continuum emission located at
R. A. $=\,20^h \,51^m \,29.86^s$,
Decl $=60\degr \,18\arcmin \,38\farcs{23}$ in J2000 (\S\ref{ss:cont}).
The dashed lines from the star indicate the major axes of the red lobes
(see Table \ref{tbl:outflow} for the P.A.).
All the other symbols are the same as in Figure \ref{fig:lobemaps}.
\label{fig:redlobemaps}}
\end{figure}

We therefore made another set of outflow lobe maps 
shown in \textcolor{black}{Figure \ref{fig:redlobemaps}}
which presents a comparison of 
the redshifted intermediate-velocity emission (hereafter ``red IV")
integrated over
$+0.2\leq v_\mathrm{LSR}/\mathrm{km\,s^{-1}}\leq +5.4$ and
the redshifted high-velocity emission (hereafter ``red HV") over
$+5.4< v_\mathrm{LSR}/\mathrm{km\,s^{-1}}\leq +7.0$.
These velocity ranges are shown by the horizontal color bars in Figure \ref{fig:sp}.
Note that the ``red IV" is  composed of 
the main emission associated with the continuum source and 
the isolated emission seen to the southwest (see Figure \ref{fig:redlobemaps}a).
For the ``red IV", we consider that our observations were not sensitive enough to 
detect a structure possibly 
connecting the mainbody and the southwestern island.\par

\textcolor{black}{Last, we point out \textcolor{black}{an observational fact}
that the boundary velocity between the ``red IV" and ``red HV" lobes is
close to the LSR-velocity of the \wat\ maser emission at 22\,GHz 
\citep{rsf03}(see Figure \ref{fig:sp}).}
The co-existence of the prominent CO wing emission and the maser transition
which is widely believed to be excited in a shocked region 
\citep[e.g., ][]{Hollenbach93, Hollenbach13, rsf01},
strongly suggests that 
\textcolor{black}{the masers are collisionally excited 
in the shocked region between the red-IV lobe with the ambient medium
(see Figure \ref{fig:sp}).}\par

In summary, 
\textcolor{black}{
our SMA observations clearly detected a prominent 
high-velocity outflow in the CO (3--2) emission towards the
central object in the core.
Because no other continuum sources are detected, 
the continuum source must be driving the outflow.}
The compact outflowing gas has the following features:
(i) both the blue and red lobes are bright at the southwest of 
the continuum source,
(ii) bipolarity between the blue- and red lobes cannot be uniquely identified 
with respect to the continuum source position, 
(iii) the blue lobe exhibits an elongated structure 
along the northeast-southwest direction, and
(iv) 
\textcolor{black}{
the red lobe can be resolved into the dual structures of 
the ``red IV" lobe and the ``red HV" one.
The ``red IV" lobe has a wide opening angle,}
\textcolor{black}{
whereas the ``red HV" lobe is more collimated than the ``red IV":
the former apparently traces the central axis of the latter.
}
\par

\subsubsection{Physical Properties of the Compact Outflow Lobes}
\label{sss:outflow}
Although the spatial relationship between the outflow lobes and the driving source is puzzling, 
we derived the physical properties of the lobes: 
size, position angle (P.A.), opening angle ($\theta$), flow velocity, 
dynamical time scale ($\tau_\mathrm{dyn}$),
mass ($M_\mathrm{lobe}$), 
mass loss rate ($\dot{M}_\mathrm{outflow}$), and 
momentum rate ($F_\mathrm{outflow}$).
These parameters are summarized in Table \ref{tbl:outflow}.\par

\textcolor{black}{
For the red lobes, we calculated the outflow properties 
for the ``red IV" and ``red HV" ones (see Figure \ref{fig:redlobemaps}).
We spatially divided the blue lobe into the
northeastern and southwestern ones (see Table \ref{tbl:outflow}).
The two blue lobes are separated by the line of
P.A.$=-45\degr$ which passes the peak of the continuum emission.
Notice that the two blue lobes are defined over the common LSR-velocity range
(\S\ref{sss:velstructure}).
}
Adopting \textcolor{black}{
the axis lines which pass the continuum peak and have the position angles in Table \ref{tbl:outflow}
as reference axes,
}
we measured the opening angle by eye because 
it is not easy to uniquely define the lobe shapes.
We estimate that the uncertainties in $\theta$ would 
be $\Delta\theta \sim$ 30\degr.\par


The dynamical time scale $\tau_\mathrm{dyn}$ is
estimated from the ratio between the maximum extent of the lobe and 
the flow velocity which is given by $v_\mathrm{flow}\,=\,|v_\mathrm{t} - v_\mathrm{sys}|$ 
(\S\ref{sss:spectra}.
\textcolor{black}{
; Table \ref{tbl:outflow}).
}
Moreover, because 
\textcolor{black}{
no information about the outflow inclination $i$ is available from observations, 
we assumed an outflow inclination angle ($i$) of $45\degr$.
\textcolor{black}{
Here we defined $i$ by the angle between the outflow axis and the line of sight.
}
In order to see the uncertainty caused by the unknown $i$ values,
we present \textcolor{black}{Figure \ref{fig:InclAngles}} 
which shows how outflow velocity 
and $\tau_\mathrm{dyn}$ change with outflow inclination angles.} 
We obtained the $\tau_\mathrm{dyn}$ range
from 500 to \textcolor{black}{2000 yrs} for the ``red HV" and ``red IV" lobes by
\textcolor{black}{
correcting for $i = 45\degr$.
}
Here we did not use \Vch\ calculated in \S\ref{sss:velstructure}
for the $\tau_\mathrm{dyn}$ estimates because
the use of \Vch\ overestimates the $\tau_\mathrm{dyn}$ values, 
\textcolor{black}{which propagate to the other physical properties.}
\par

\textcolor{black}{
In \S\ref{sss:spectra}, we described that  
the outflow wing emission is detected only towards the center position
among all the CSO spectra}
\textcolor{black}{(see the \Vlsr\ $= 9.2$ \kms\ component in the 
central panel of Figure \ref{fig:pmap}).
Assuming that the 9.2 \kms\ component represents line-of-sight velocity 
of a redshifted lobe, 
we can have another inference of $\tau_\mathrm{dyn}$:
the beam size for the single-dish \co\ (3--2) observations and the flow velocity 
given by 
$v_\mathrm{flow} = |v_\mathrm{LSR} - v_\mathrm{sys}|/\cos(45\degr) \simeq 17$ \kms\
yields an upper limit of 
$\tau_\mathrm{dyn}  \lesssim l/v_\mathrm{flow}=$\,900\,yrs
where 
$l \lesssim (\theta_\mathrm{HPBW}/2)/\sin(45\degr) =$\,3100\,AU.
The upper limit does not contradict with
the range of $\tau_\mathrm{dyn} \simeq$ 500 -- 2000 yrs 
estimated from the interferometric images.
}

For calculating $M_\mathrm{lobe}$ we assumed that 
the \CO\ wing emission is optically thin and its excitation is in LTE.
\textcolor{black}{ 
We keep using the \co/H$_2$ abundance ratio of $10^{-4}$ \citep{Dickman78}
for a comparison with previous papers.
We adopted the range of the excitation temperature to be between 7.3\,K and 22.6\,K.
Here, the lower limit is given by 
adding the temperature of 
the cosmic microwave background emission to
the peak brightness temperature of the CO spectrum (Figure \ref{fig:sp}), 
whereas the upper limit is the gas temperature described in \S\ref{ss:cont}.
We also measured mean \Tex\ of the gas over the regions enclosed by the 3\sgm\ contours 
of the blue- and redshifted lobes to be 
20.4$\pm$3.0\,K and 17.4$\pm$5.6\,K, respectively (Appendix \ref{as:Tex}),
both of which fall in the range.
Because it is not necessarily certain whether or not the \Tex\ of the N$_2$H$^+$ (1--0) lines,
which generally probe a static dense core gas,
traces the temperatures of the outflowing gas where the CO transitions are excited,
it is reasonable to adopt such a \Tex\ range rather than using a single value.
}
\par

By integrating the lobe emission shown in Figure \ref{fig:lobemaps},
we estimated $M_\mathrm{lobe}$ to be of the order of
$10^{-5}-10^{-3}$ \Msun. The lobe masses would be underestimated
for the following two reasons.
One is that the lobe emission was assumed to be optically thin. 
The other is the uncertainty in defining the boundary velocities
between the outflowing gas and the ambient quiescent gas
due to the effect of the resolve-out and/or the self-absorption 
(\S\ref{sss:velstructure}).
Both the northeastern and southwestern blue lobes
have comparable $\tau_\mathrm{dyn}$ and $M_\mathrm{lobe}$ values.
\textcolor{black}{Notice that the majority of the red lobe mass
is attributed to the ``red IV" one.}\par

\textcolor{black}{
After obtaining $M_\mathrm{lobe}$, 
the mass-loss, momentum rates, and mechanical luminosity are estimated to be 
$\dot{M}_\mathrm{outlow} \sim 10^{-9}-10^{-6}$ \Msun\ yr$^{-1}$,
$\dot{F}_\mathrm{outlow} \sim 10^{-8}-10^{-6}$ \Msun\ \kms\ yr$^{-1}$, and
$\dot{L}_\mathrm{outlow} \sim 10^{-5}-10^{-2}$ \Lsun, respectively (Table \ref{tbl:outflow}).
}
The ``red IV" lobe is one order of magnitude more powerful than the blue lobes
(see $F_\mathrm{outflow}$ in Table \ref{tbl:outflow}).
In \S\ref{sss:velstructure} we pointed out that the spatial extent of the ``red HV" lobe
corresponds to the central axis of the ``red IV" lobe.
However, a comparison of their $F_\mathrm{outflow}$ values suggests that the
``red HV" lobe is too powerless to drive the ``red IV" lobe.\par

We \textcolor{black}{argue} that the outflow lobes in the \gf\ core is
clearly one of the smallest, the least massive, and 
the least powerful outflows
when we compare with other sources compiled 
in e.g., \citet{Beuther02, Takahashi08, Takahashi12, Bally16}.

\section{Discussion}
\label{s:discussion}

\subsection{Excluding ``Class 0 proto-brown dwarf" and Wide Binary Scenarios}
\label{ss:singleStar}
Considering an upper limit of the luminosity reported in the previous work
\citep[\Lbol $< 0.3$; ][]{Wiesemeyer97}, 
\citet{Palau14} categorized this object as a Very Low Luminosity Objects 
\citep[VeLLOs; see][for definitions]{Kauffmann05, diFrancesco07}
more specifically, a ``class\,0 proto-brown dwarf".
However, this interpretation  
\textcolor{black}{may be unlikely
because of the ``envelope infall rate" with the order of 
$10^{-5}$ \Msun\ yr$^{-1}$ (\S\ref{ss:StellarMass})
which is certainly larger than the upper envelope of 
$\dot{M}_\mathrm{acc}\sim 10^{-9} - 10^{-6}$ \Msun\ yr$^{-1}$ 
for VeLLOs discussed in e.g., \citet[][]{Pineda11}.
It should be noticed that this comparison should be made among infall rates
measured in the core-scale at 0.1--0.01\,pc, not in the envelope-disk scale 
at $\lesssim 10^2$\,AU,
because majority of the previous works used data obtained with the core-scale.
} 
\par

\textcolor{black}{In the previous work we pointed out that there is a positional
offset of 6\arcsec\ between
the 3\,mm source and the peak of the \NtwoH\ (1--0) emission (see Figure 15 in Paper I),
and proposed that the offset
can be interpreted as the presence of a binary system.
Clearly such a wide binary scenario is rejected by the higher resolution
SMA observations.}
However, we do have neither positive evidence to support the presence of 
a closer binary whose separation is smaller than our beam sizes ($\lesssim 400$\,AU)
nor negative one to reject such a possibility.
\textcolor{black}{
Therefore, we keep our assumption that the outflow is driven by a single object
to make our discussion as simple as possible.}

\subsection{Age of the Outflow Driving Source}
\label{ss:age}
\textcolor{black}{It should be noticed that the 
range of dynamical time scale of the outflow lobes
(500 -- \textcolor{black}{2000 yrs}; \S\ref{sss:outflow}; Table \ref{tbl:outflow})
is comparable or} shorter than 
those for typical class 0 sources \citep[$10^3$ -- $10^4$ yrs; e.g.,][]{Bachiller96, Bontemps96, Arce06, Curtis10, Velu14}.
Contrary to the red lobes, the blue lobes seem to be a few times older than 
the red ones, but their $\tau_\mathrm{dyn}$ values are still
comparable to those of the youngest class 0 sources.\par

\textcolor{black}{Figure \ref{fig:spatialcomp}} demonstrates the compactness of the 
outflow system by 
a comparison of the size scales between the 0.1\,pc-scale molecular cloud core and
\textcolor{black}{the 1000 AU-scale circumstellar structure, i.e.,} envelope.
\textcolor{black}{Here the cloud core is represented by the \HtCOp\ (1--0) line
whereas the envelope by the 350 \micron\ continuum emission} (Paper I).
The total extent of the outflow is about $1/40$ of that of the cloud core
and is about $1/5$ of that of the  envelope when measured along its polar direction (NE-SW),
indicating that the central object must be in its early evolutionary stage.
Moreover the compactness of the outflow indicates that 
the outflow has not yet dispersed the parental core.
The overall NE-SW direction of the outflow system
is almost perpendicular to the elongation of the 350 \micron\ envelope,
suggestive of on-going disk-mediated accretion.
This orthogonality is not affected by the absolute position accuracy 
of the 350 \micron\ image ($\lesssim 4\arcsec$; Paper I).
Hence, if an edge-on disk exists, 
its elongation should be almost parallel to that of the envelope. \par

\textcolor{black}{In addition to the outflow dynamical time scale,
we can obtain another constraint on the source age from the size of 
the free-fall region centered on a forming protostar.}
\textcolor{black}{Despite the presence of the compact outflow,}
\textcolor{black}{in the radial column density profile, $N(r)$ (Figure 11 in Paper I), 
we identified that the best-fit power-law profile with an index of $-1$ 
holds down to $\lesssim 600$ AU and a free-fall profile with 
an index of $-1/2$ was not recognized in $r\gtrsim 600$ AU.}
\textcolor{black}{Because the gas motions over the core is well described by 
the extended Larson-Penston's solution for $t>$ 0,
i.e., the LPH solution (Papers I and III),
the absence of an $N(r)\propto r^{-1/2}$ profile,
i.e., $\rho(r)\propto r^{-3/2}$ in volume mass density, 
in the $r\gtrsim 600$\,AU region indicates that free-fall region 
around the protostar
has not yet expanded 
up to the radius of $\sim 600$ AU, 
yielding to an upper limit of the central free-fall region ($r_\mathrm{ff}\lesssim 600$ AU).}
\textcolor{black}{Recall that we conservatively adopted 
$r_\mathrm{ff}\lesssim 1200$ AU in Paper I 
because of the 6\arcsec\ separation-binary interpretation (\S\ref{ss:singleStar}).
We now use 
$r_\mathrm{ff}\lesssim 600$\,AU and $T = 20$\,K (\S\ref{sss:outflow}),
instead of $r_\mathrm{ff}\lesssim 1200$\,AU and $T = 10$\,K used in Paper I,
the upper limit of the elapsed time since the first kernel of a mass has formed is updated as,
}
\begin{eqnarray}
\tau_\mathrm{*}\, 
& = & \frac{r_\mathrm{ff}}{~(2\sim 3)c_\mathrm{iso}~} \label{eqn:tau_LPH}\\
& \lesssim\, & 
\textcolor{black}{
(4\pm 1)\times 10^3\,\mathrm{yrs}
}
\left(\frac{T}{20 \,\mathrm{K}}\right)^{-1/2}
\textcolor{black}{
\left(\frac{r_\mathrm{ff}}{600\,\mathrm{AU}}\right)
}.
\label{eqn:tau_protostar}
\end{eqnarray}
The revised upper limit is consistent with the 
dynamical time scales of the red and blue lobes
(see Table \ref{tbl:outflow}).
\textcolor{black}{Last, the $r_\mathrm{ff}\lesssim 600\,\mathrm{AU}$
does not contradict with the envelope inner radius
(corresponding to $R_\mathrm{max}^\mathrm{disk}$ in Table \ref{tbl:sedfitter};
described in \S\ref{ss:SED}).}
\subsection{``Stellar Mass" of the Outflow Driving Source}
\label{ss:StellarMass}
\textcolor{black}{
Next we attempt to estimate the stellar mass using the mass
infall rate in the envelope and the stellar age.}
Considering the two facts that 
the gas is \textcolor{black}{
globally ($r\lesssim 30\arcsec$, i.e., $r\lesssim$0.029 pc) 
infalling onto the central object (Paper III)}
and that the $\rho(r)\propto r^{-2}$ profile
\textcolor{black}{
is identified for $r\gtrsim 600$\,AU (\S\ref{ss:age}),
the previously estimated global mass infall rate over the core 
($\dot{M}_\mathrm{inf}^\mathrm{sph} = 2.5\times 10^{-5}$ \Msun\ yr$^{-1}$)
is considered to be valid in the region of $600 \lesssim r/[\mathrm{AU}]\lesssim 6000$.}
\textcolor{black}{Indeed, the same assumption, i.e., 
the global infall rate should represent the mass accretion rate onto a stellar core,
was adopted in \citet{Maureira17} when they discuss evolutionary phase 
of a \fc\ candidate.}\par

\textcolor{black}{
However, this spherical infall rate would not hold for the inner
$r\lesssim\,600$\,AU region. This is because the compact outflow has already launched, 
suggestive of non-spherical \textcolor{black}{disk-mediated}
accretion onto the forming star due to the dispersal of the envelope 
by the outflow (\S\ref{sss:velstructure}).
Therefore, we take a non-spherical}
\textcolor{black}{steady accretion rate of
$\dot{M}_\mathrm{acc}\,=\,f\dot{M}_\mathrm{inf}^\mathrm{sph}$ for $r\lesssim 600$\,AU
where an efficiency coefficient $f$ is written as
$(4\pi -\Omega_\mathrm{outflow})/(4\pi)$ with
a total solid angle of the blue and red lobes viewed from the central star
($\Omega_\mathrm{outflow}$).
Assuming the shapes of the bipolar outflow lobes are conical, 
we calculated the total solid angle of the lobes as
$\Omega_\mathrm{outflow} = 2\cdot 2\pi[1-\cos(\theta/2)]\sim 5.4$ steradian for the
$\theta \sim 110\degr$ which is the widest opening angle of the
``red IV" lobe (Table \ref{tbl:outflow}).
We obtained $f\sim 0.6$,
\textcolor{black}{leading to the non-spherical steady accretion rate of,}
}
\begin{equation}
\dot{M}_\mathrm{acc} = 
f\dot{M}_\mathrm{inf}^\mathrm{sph}\sim 1\times 10^{-5} \left(\frac{f}{0.6}\right) \Msun\ {\mathrm yr}^{-1}.
\label{eqn:Mdot}
\end{equation}
Then the object may have acquired a stellar mass of,
\begin{equation}
M_\ast \,=\, 
\textcolor{black}{
\dot{M}_\mathrm{acc}\,\tau_\mathrm{*}\,= f\dot{M}_\mathrm{inf}^\mathrm{sph}\,\tau_\mathrm{*}\,\lesssim\,
0.06\,\Msun\,
\left(\frac{f}{0.6}\right)
}
\left(\frac{\dot{M}_\mathrm{inf}^\mathrm{sph}}{\,2.5\times 10^{-5}\,\Msun\ \mathrm{yr}^{-1}\,}\right)
\left(\frac{\tau_\mathrm{*}}{\,4\times 10^3\,\mathrm{yrs}\,}\right),
\label{eqn:Mstar}
\end{equation}
since \textcolor{black}{
the formation of the central point source.}
Here the upper limit in Eq.(\ref{eqn:Mstar})
is due to the upper limit in the $\tau_\ast$ estimate
[Eq.(\ref{eqn:tau_protostar})].
\textcolor{black}{Notice that the $M_\ast$ value does not change significantly 
even if one consider the range of the 
outflow dynamical time scales, $\tau_\mathrm{dyn}$, as the age of the object.}\par

\textcolor{black}{
We must keep in mind that the current accretion/outflow rates may 
much differ from the time-averaged rates during the formation process
of the possible disk-outflow system or a possibility that
an accretion rate from the envelope onto a putative disk should differ from
an accretion rate from the disk onto the central object.
However, because such more rigorous discussion is beyond the scope of
this paper, we assume that the accretion rates onto
the disk and the forming star are equal to each other and steady for further discussion.}\par


%
%
%

\subsection{Spectral Energy Distribution of the Outflow Driving Source}
\label{ss:SED}
In order to obtain further constraints on the nature of the source,
we produced \textcolor{black}{Figure \ref{fig:sedfitter}} where
spectral energy distribution (SED) of the outflow driving source between the mm bands
and Spitzer bands is presented.
\textcolor{black}{The model SEDs were obtained by using a \texttt{sedfitter} tool 
by \citet{Robitaille06} together with 
large sets of SED models provided by \citet{Robitaille17}.
As emphasized by the author, we have to keep in mind the limitations of the tool
and the model sets
because they provide simplified SED model sets to search for 
a first-order-of-approximation picture of a YSO of interests among the various
combinations of physical parameters characterizing a YSO.}
Each model set in \citet{Robitaille17}
allows us to search for the best-fit solution in a wider range of the model parameters 
than the original one \citep{Robitaille06}.
Each set consists of a protostar or a pre-main sequence (PMS) star surrounded by an 
accretion disk and a rotationally infalling envelope with a bipolar outflow,
and also considers the scattering and reprocessing of the stellar radiation by dust. 
Moreover, some model sets include a bipolar cavity dispersed by an outflow. 
In each model set, 
model SEDs were computed at 10 different viewing angles by convolving 
a frequency-filter function
and an aperture size at each measurement; we used the default ones for the
Spitzer measurements, and added those for our CSO, SMA, CARMA, and OVRO observations.\par


To fit the photometry results (Tables \ref{tbl:phot} and \ref{tbl:apphot}), 
we set a distance range of 200$\pm$20\,pc,
and used an extinction curve by \citet{Whitney03, Whitney04},
allowing to estimate the foreground $A_v$ as well
(see the $A_v$ value in each panel of Figure \ref{fig:sedfitter}). 
We dealt with the 70 \mic\ flux density as an upper limit
because the HPBW of the aperture encompassing the source 
is much larger than those imaged by SMA, CARMA and OVRO interferometers.
Notice that the number of the free parameters in the model sets
\citep[see Table 2 in][]{Robitaille17} ranges between 2 and 12, 
whereas we have \textcolor{black}{fifteen data points ($n_\mathrm{data} = 15$)}, 
including the upper limits.
A caution must be used for \textcolor{black}{the WISE 22 \mic\ and Spitzer 24 \mic\
data whose aperture sizes are} comparable
to that for the 350 \mic\ (Table \ref{tbl:apphot}).
\textcolor{black}{This is because}
we did not consider them as upper limits 
because the \textcolor{black}{22 and 24 \mic\ emission} most likely comes from 
the central compact component(s) rather than from the extended envelope 
traced by the 70 and 350 \mic\ emission.
Indeed, we found that all the models gave us unrealistic SED fits
when we set \textcolor{black}{the 22 and 24 \mic\ fluxes as upper limits.}\par

As mentioned above, we attempted all the model sets summarized in Table 2 of \citet{Robitaille17},
and selected the \textcolor{black}{most reasonable models from each model set}
based on the criterion of 
$\chi^2 - \chi_\mathrm{best}^2 < 5 \,n_\mathrm{data}$.
In practice, we rejected \textcolor{black}{the 16 out of 18} model sets in Table 2 of \citet{Robitaille17}
because these model sets failed to reproduce the observed SEDs,
\textcolor{black}{whereas the remaining two model sets of
\texttt{spu-smi} and \texttt{spubhmi}}
gave reasonable fits.
\textcolor{black}{Here the first character of \texttt{s} in the model-set names denotes that they 
commonly consider emission from a central \texttt{s}tar, 
the second one of \texttt{p} does \texttt{p}assive disk,
the third \texttt{u} does so-called Ulrich-like envelope \citep[an envelope with radial power-law 
dependency of $\sim 3/2$ outside the centrifugal
radius $R_c$, 1/2 inside;][]{Ulrich76}.
The fourth \texttt{b} in \texttt{spubhmi} indicates that this model set
considers a \texttt{b}ipolar-outflow cavity, whereas not in \texttt{spu-smi}.
The fifth \texttt{h} indicates that the \texttt{spubhmi} model set
leaves radius of the inner hole produced by the outflow as a free-parameter, 
whereas the fifth \texttt{s} means that the \texttt{spu-smi} adopted
a dust-sublimation radius as an inner hole radius.
The last two characters of \texttt{mi} indicates that 
the both model sets commonly considered the ambient medium \texttt{(m)} together with
the interstellar dusts \texttt{(i)}, as described earlier.}\par

\textcolor{black}{On the basis of the inferred physical parameters, 
we rejected \texttt{spu-smi}
because the stellar radius suggested by the model sets
($R_\ast = 0.12$ \Rsun) is too small and
the stellar surface temperature ($T_\ast \sim 1.9\times 10^4$\,K) 
is too high for a protostar (\S\ref{ss:age}).}
\par


\textcolor{black}{The remaining model sets of \texttt{spubhmi} yields the decent SED fit
(Figure \ref{fig:sedfitter}) with the reasonable physical parameters (Table \ref{tbl:sedfitter}).
The model gave plausible values of $R_\ast \sim 4$ \Rsun,
and $T_\ast \sim 3\times 10^3$\,K as well as those characterizing
the envelope and disk.
The inferred inclination angle of $i\sim 65\degr$ from the two models
gives a constraint in our analysis in \S\ref{sss:outflow}, and will be discussed
in \S\ref{ss:EvolutionFromOutflow}.
In addition, the suggested cavity opening angle of $2\theta\sim 40\degr$
may not contradict with our estimate in Table \ref{tbl:outflow}
when we consider the uncertainty of our measurement
($\Delta\theta\sim 30\degr$; \S\ref{sss:outflow}).
Last, the inferred $R_\mathrm{disk}$ value
does not affect the estimate of the protostar's age based on the radial
density profile (\S\ref{ss:age}).}\par

\textcolor{black}{In conclusion, 
it is highly likely that we are dealing with 
an edge-on outflow system
driven by a $T_\ast \sim 3000$\,K protostar
surrounded by a $\sim 200$ AU radius disk with an infalling envelope.
Moreover, we argue that 
the outflow system whose lobe axis is almost parallel 
to the plane of sky may not have completed dispersing its natal cloud core.}

\subsection{Evolutionary Stage of the Outflow Driving Source: Constraints from the Outflow Properties}
\label{ss:EvolutionFromOutflow}
\textcolor{black}{The SED analysis suggested that 
the outflow is viewed almost from its side
\citep[see e.g., Figure 7 in][]{TT11}.
In this subsection, we discuss this geometry along with 
the outflow properties (\S\ref{sss:outflow}).}

\subsubsection{Terminology of the Outflow Velocity and Inclination Angle
Dependency on the Flow Velocity and Dynamical Timescale}
\label{sss:IncAngle_Vflow_taudyn}
Before addressing \textcolor{black}{the outflow geometries}, we must clarify 
the terminology of the outflow velocities when 
we compare to those in theoretical studies.
Most of the theoretical studies refer to ``high velocity" as those having 
$v_\mathrm{flow}$ of the order of $10^1 - 10^2$ \kms\
and ``low velocity" as 1 \kms\ $\lesssim v_\mathrm{out} \lesssim 10$ \kms\
\citep[e.g.,][]{Bachiller96}.
Note that these 
\textcolor{black}{theoretical} 
velocities are the 3D one.
In contrast, observers argue 
outflow velocities projected onto the line of sight.
Furthermore there is no widely accepted unique
definition of the velocity range for ``high" and ``low" velocities even if
we limit to discussing outflows traced by low-$J$ CO lines.
Considering these limitations and the uncertainties in our observations,
we do not directly compare observed outflow velocities and 
\textcolor{black}{
those in numerical simulations. Therefore,
} 
we keep using
the words of the IV and HV defined in \S\ref{sss:velstructure}, 
and use ``lower velocity" and ``higher velocity" in a relative sense.\par

\textcolor{black}{As already seen in Figure \ref{fig:InclAngles},
3D flow velocity of $v_\mathrm{flow}/\cos i$ and the ``real"
dynamical time scale of $\tau_\mathrm{dyn}\cos i/\sin i$ 
for the four lobes 
strongly depends on $i$.
In addition, the opening angles of the lobes are rather large of
$\theta \sim 70\degr - 110\degr$ with the uncertainties of 
$\Delta\theta\sim 30\degr$ (Table \ref{tbl:outflow}). 
We therefore discuss two representative cases of 
the \textcolor{black}{pole-on outflow at $i = 20\degr$ and the edge-on one at $i = 70\degr$,
as done in e.g., \citet[][]{Belloche06}.}
}\par

\subsubsection{``Pole-on" vs. ``Edge-on" Outflows Based on the Outflow Age and the Elapsed Time 
of Mass Accretion}
\label{sss:Geometry_by_Ages}
\begin{figure}
\includegraphics[angle=0,scale=.4, bb=0 0 18cm 7.8cm]{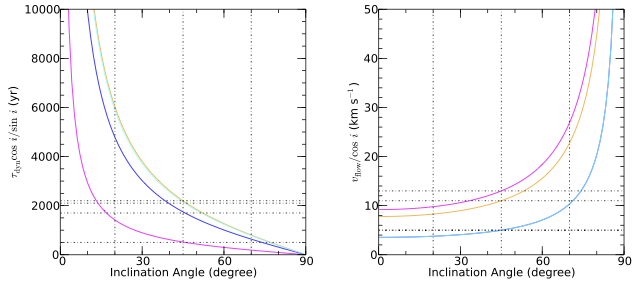}
\caption{Plots of the outflow velocity ($v_\mathrm{flow}/\cos\,i$; the $left$ panel) 
and dynamical time scale ($\tau_\mathrm{dyn}\cos i/\sin i$; the $right$ panel) 
of the outflow lobes 
as a function of the inclination angle ($i$). 
Here $i$ is defined by the angle measured from the line-of-sight (l.o.s.; $i = 0\arcdeg$), 
hence a lobe with $i = 90\degr$ is parallel to the sky plane.
\textcolor{black}{We refer to an outflow whose axis has $i = 0\arcdeg$ as a pole-on outflow, 
whereas $i = 90\arcdeg$ as an edge-on outflow (\S\ref{sss:IncAngle_Vflow_taudyn}).
The magenta, orange, cyan, and blue curves correspond to the
inclination dependence of the ``Red HV", ``Red IV", ``Blue NE", and ``Blue SW" lobes, respectively.
See \S\ref{sss:velstructure} for the definition of each lobe.
The $v_\mathrm{flow}/\cos i$ and $\tau_\mathrm{dyn}\cos i/\sin i$ curves in these plots 
pass the values in Table \ref{tbl:outflow} at $i = 45\arcdeg$.
Note that the cyan and blue curves in the right panel agree with each other.}
\label{fig:InclAngles}}
\end{figure}

\begin{figure}
\includegraphics[angle=0,scale=.34, bb=0 0 18cm 20.2cm]{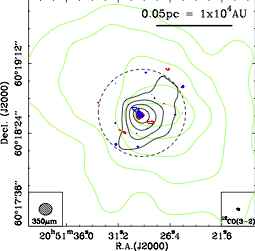}
\caption{
Comparison among the dense molecular cloud core traced by the
\HtCOp\ (1--0) emission (green contour), 
the circumstellar envelope observed by the 350 \micron\ continuum emission 
(black contours)
and the newly detected compact molecular outflow (the blue- and red contours; 
\textcolor{black}{see Figure \ref{fig:lobemaps}}).
The \HtCOp\ line and  350 \micron\ continuum emission maps are taken from Paper I.
All the contours are plotted by the $3\sigma$ intervals 
starting from the $3\sigma$ levels.
The ellipses in the bottom-left and bottom-right corners are the beam size
of the SHARCII bolometer for the 350 \micron\ imaging and the
synthesized beam size for the SMA \co\ (3--2) line observations, respectively.
The dashed circle at the center indicates the field-of-view of the SMA \CO\ (3--2) observations
(see Table \ref{tbl:obs}).
\label{fig:spatialcomp}}

\end{figure}

\begin{figure}
\includegraphics[angle=0,scale=.48, bb=0 0 15cm 3.8cm]{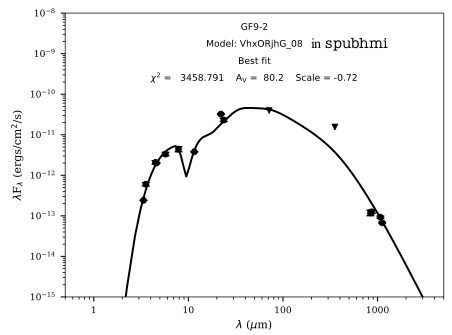}
\caption{Model fitting to the \gf\ protostar
performed by \texttt{sedfitter} tool using different YSO SED models in \citet{Robitaille17}. 
\textcolor{black}{The solid black line shows the best-fit model.}
See Tables \ref{tbl:phot} and \ref{tbl:apphot} for the flux densities, Table \ref{tbl:sedfitter} for the inferred parameters and, \S\ref{ss:SED} for details. 
\label{fig:sedfitter}}
\end{figure}

\paragraph{\textcolor{black}{A Pole-on Outflow (e.g., $i = 20\degr$)}}
Figure \ref{fig:InclAngles} suggests that the 3D-velocities of the 
\textcolor{black}{pole-on} 
lobes range from 3 to 9 \kms, leading to the corrected ages of
$\tau_\mathrm{dyn}\cos i/\sin i \sim 10^3 - 10^4$ yrs.
In the \textcolor{black}{pole-on geometry}
the large spatial-overlapping between 
the blue and red lobes together with the absence of the redshifted CO emission 
to the northeast of the continuum source
\textcolor{black}{is} explained to some extent.
This is because the outflow gas moving away from us along the 
\textcolor{black}{outflow} 
axis \textcolor{black}{should} be obscured by the circumstellar materials.\par

However, there are two major caveats in \textcolor{black}{the pole-on} outflow hypothesis.
If the \textcolor{black}{pole-on} is the case, all the lobes should show more roundish morphology
\textcolor{black}{to form a symmetric roundish lobe pairs,}
but the ``red HV" and blue lobes are elongated.
It is also impossible to explain the southwest island (see Figure \ref{fig:lobemaps}b)
along with the \textcolor{black}{pole-on} hypothesis.
\textcolor{black}{
The other major caveat is that the $i$-corrected 
$\tau_\mathrm{dyn}$ values of the order of $10^3 - 10^4$ yrs 
would be larger than the source age of 
$\tau_* \lesssim 4\times 10^3$ yrs in Eq.(\ref{eqn:tau_protostar}).}
\textcolor{black}{We therefore conclude that a pole-on geometry is unlikely.}
\par

\paragraph{\textcolor{black}{An Edge-on Outflow (e.g., $i =$ 70$\degr$)}}
If we are observing \textcolor{black}{an edge-on outflow}, 
the 3D-velocities of the lobes would be approximately 10 -- 30 \kms\ (Figure \ref{fig:InclAngles}),
leading to the lobe ages of 
$\tau_\mathrm{dyn}\cos i/\sin i \lesssim 1\times 10^3$ yrs.
\textcolor{black}{It is well-known that protostars, i.e., class 0 sources,
show highly-collimated high-velocity outflows whose dynamical time scale
ranges $10^3$ -- $10^4$ yrs \citep[e.g.,][]{Gueth99, Santangelo15}.
The inferred age range of the \gf\ outflow agrees with the range, 
and is considered to represent} elapsed time of 
the ongoing accretion process (\S\ref{ss:age}).
If such \textcolor{black}{an edge-on} geometry is the case, 
all the caveats raised in the \textcolor{black}{pole-on hypothesis}, 
except for the absence of distinct bipolarity, may be resolved.
In addition, the spatial overlapping of the blue and red lobes seen to the southwest of the continuum source 
is also reasonably explained by 
\textcolor{black}{edge-on outflow lobes with large opening angles}
of $\theta \sim 70\degr - 110\degr$ (Table \ref{tbl:outflow}).\par

However, the \textcolor{black}{edge-on} hypothesis does not explain why the morphology and 
velocity structure of the blue and red lobes are not symmetric; 
we should detect redshifted lobe(s) to the northeast of the continuum source.
\textcolor{black}{
Such asymmetry may be explained by
``realistic" MHD simulations which includes turbulence \cite[e.g.,][]{MH11}.
}\par

Keeping all the strong and weak points in mind,
it is possible that the central object has experienced its second collapse
$\tau_\mathrm{dyn}\cos i/\sin i \lesssim 1\times 10^3$ yrs ago.
In this interpretation, 
the poorly collimated ``red IV" lobe (Figure \ref{fig:redlobemaps})
would represent a remnant outflowing gas which is the fossil of a lobe that had been driven
by the object when it had been at the \fc\ stage, and
the ``red HV" lobe 
(Figure \ref{fig:redlobemaps}) is considered as a fresh lobe currently 
being driven by the protostar.
If this is the case, the dynamical time scale of the ``red IV" lobe
($\tau_\mathrm{dyn}\cos i/\sin i \sim $500 yrs for $i\sim 70\degr$; Figure \ref{fig:InclAngles})
may be interpreted as the duration of the \fc\ stage.
\textcolor{black}{In this interpretation, the small $\beta$ of 0.4$\pm$0.3 (\S\ref{sss:ContSpIndex})
should be attributed to a remnant of the first hydrostatic core.}
\par

\subsubsection{Summary of the Subsection}
\label{sss:OutflowSummary}
Considering the geometry of the compact outflow lobes in conjunction with the
$\tau_\mathrm{dyn}$ and $\tau_\ast$ estimates,
we suggest that the outflow geometry is 
\textcolor{black}{edge-on 
which reconciles with the results from the SED analysis.}
\section{Summary}
\label{s:summary}
Interferometric observations of the \CO\ (3--2) line and
the continuum emission
at 3.3\,mm, 1.1\,mm, and 850 \micron\ bands were carried out 
towards the deeply embedded 
protostar at the center of the dense molecular cloud core \gf\
using CARMA and SMA.
\textcolor{black}{
Furthermore we analyzed the $Spitzer$ and $WISE$ satellite images} and
the single-dish \CO\ (3--2) spectra previously taken with CSO.
The main findings of this research are summarized as follows.

\begin{enumerate}
\item At the center of the cloud core, 
we detected a compact continuum source which 
is considered to be representing a circumstellar envelope 
\textcolor{black}{at 1.1\,mm and 850 \micron\ bands.}
The beam-deconvolved effective radius, \Reff, of the continuum source 
was measured to be \ReffValueUnit\ at the 1.1\,mm and 850 \micron\ bands where
we attained the linear resolution of $\sim 400$\,AU 
at the distance of the source (200\,pc).
\label{enu:phot}

\item \textcolor{black}{Towards the position of the submm source, 
an infrared source, 
\textcolor{black}{designated as WISE J205129.83+601838,} 
is clearly detected at wavelengths between 70 \micron\ and \textcolor{black}{3.4} \micron\
in the Spitzer MIPS and IRAC images \textcolor{black}{as well as WISE ones}.
This detection clearly indicates that the object is at} a protostar phase.
Our SED analysis using the \texttt{sedfitter} tool \citep{Robitaille17}
suggested its surface temperature of \textcolor{black}{$\sim 3000$\,K and stellar radius of $\sim$\,4 \Rsun.}

\item Our spectroscopic imaging of the \CO\ (3--2) line clearly detected 
a 1000-AU scale molecular outflow system
driven by the continuum source.
The system exhibits collimated- and poorly collimated redshifted 
outflow lobes with the line-of-sight
velocities of $\sim10$ \kms\ and $\sim 5$ \kms, respectively.
The blueshifted exhibits a collimated lobe 
extending both towards the northeast and southwest of the central source.
Our analysis showed that the outflow lobes are one of the 
youngest (dynamical time scales of $\sim$ 500 -- 2000 yrs)
and the least powerful (momentum rates of $\sim 10^{-8}-10^{-6}$ \Msun\ \kms\ yr$^{-1}$) 
ones so far detected.
A comparison between the continuum and outflow maps suggests that the innermost part 
($r\lesssim$\,2000 AU)
of the envelope is being dissipated by the outflow.

\item The momentum rate of the collimated red lobe does 
not suffice to drive the poorly-collimated red one, 
suggesting that they are $independently$ driven.
\textcolor{black}{Based on the outflow morphology, velocity structure of the lobes, 
comparisons with outflow models and the results from the SED analysis,}
we concluded that the outflow axis is \textcolor{black}{not far from} parallel to the
plane of the sky, i.e., the \textcolor{black}{edge-on} geometry.


\item \textcolor{black}{Although the outflow properties agree with 
those measured in the VeLLOs
and the ``Class 0 proto-brown dwarfs",
we excluded such interpretations on the basis of the large spherical infall rate
of the order of $10^{-5}$ \Msun\ yr$^{-1}$.
We also excluded the wide-binary interpretation based on the previous lower
resolution data
because only a unresolved ($\lesssim$400\,AU) continuum emission was detected 
by the interferometric observations.}

\item We argued that it has not passed 
\textcolor{black}{
$\tau_\ast \lesssim (4\pm 1)\times 10^3$ years}
since the protostar formed at the center of the cloud core.
This is because a radial volume density profile with a form of \RadpInf, 
which proves the presence of a freely falling gas towards the
central object, was not identified outside the 
\textcolor{black}{
$r\simeq$ 600\,AU region in radius which has a consistency
with the disk radius inferred from the SED analysis.}
The upper limit of the \textcolor{black}{protostellar} age suggests that
the total mass accreted onto the central object is
\textcolor{black}{
$M_\ast \lesssim 0.06$ \Msun.}

\end{enumerate}

\textcolor{black}{
Given the uniqueness of the source properties, 
follow-up high-resolution and high-sensitivity observations along with simulation studies are
required towards a more complete understanding of the physics in 
a low-mass protostar formation process.
}

\acknowledgments
The authors sincerely acknowledge the anonymous referee whose
comments significantly helped to improve quality of this paper, 
especially critical comments on the Spitzer data.
R. S. F. gratefully acknowledges John M. Carpenter and Andrea Isella
for their generous help and discussion at the CARMA observations 
and the data reduction process.
R. S. F. also sincerely thanks Masahiro N. Machida,
Tomoyuki Hanawa and Shu-ichiro Inutsuka for fruitful discussion, and
Takeshi Inagaki for data analysis with Python.
This work was partially supported by 
the JSPS Institutional Program for Young Researcher Overseas Visits 
({\it Wakate Haken})
at Subaru Telescope of National Astronomical Observatory of Japan (NAOJ) for R. S. F. and H. S.
and the AWA support program at Tokushima University for R. S. F.
Data analyses in this work were partly 
carried out on the computer system operated by
Subaru Telescope and that by Astronomy Data Center of NAOJ. 
The authors gratefully acknowledge all the staff at 
CARMA, SMA, CSO and the Spitzer Science Center, and 
the MIR software group at CfA, 
the AIPS software group at NRAO and
the GILDAS software group at IRAM.

%
%
\vspace{5mm}
\facilities{SMA, CARMA, OVRO mm-array, Spitzer Science Telescope, CSO 10.4\,m telescope}





\appendix
\section{Excitation Temperature Map of the \NtwoH\ (1--0) Emission}
\label{as:Tex}

As reported in Paper I, we combined the visibility data of the \NtwoH\ line 
taken with the OVRO mm-array and the single-dish Nobeyama 45\,m telescope 
(see \S 3.2.3 and Appendix A in Paper I).
Using the the combined data we performed the hyperfine
structure analysis of the \NtwoH\ transition 
(see \S 4.1. and Appendix B.1 in Paper I).
The usage of the hyperfine structure lines has an advantage that one
can assess both the \Tex\ and optical depth of the lines
with relatively high accuracy. 
In addition, such a combined image allows us to 
analyze the spatial structure of the gas with an angular resolution
that can be achieved by interferometers and the analysis is free from the ``missing flux"
problem. \par

\begin{figure}
\includegraphics[angle=0,scale=.45, bb=0 0 18cm 14.8cm]{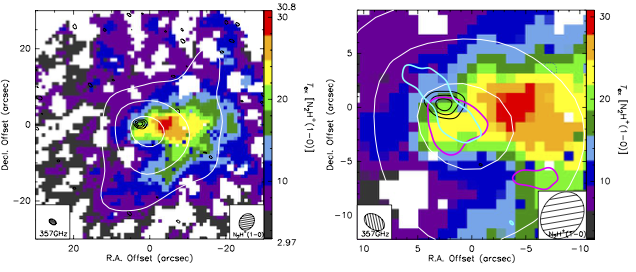}
\caption{
\textcolor{black}{(a)(left)} Excitation temperature (\Tex) map of the \NtwoH\ (1--0) line (color), 
where the 350 \micron\ (white contours; see Paper I) and 357\,GHz 
(black contours; see Figure \ref{fig:contmaps}e in this paper) 
continuum emission maps are overlaid.  
The color bar on the right-hand side shows the temperature scale in K.
The \Tex\ map was obtained from our analysis of the \NtwoH\ hyperfine structure 
in the combined image of the OVRO mm-array and single-dish Nobeyama 45\,m 
telescope data, which is free from missing flux densities.
Note that the \Tex\ map was obtained in Paper I, but was not presented there.
The contour parameters are the same as in Figure \ref{fig:spatialcomp} for the 
350 \micron\ emission and in Figure \ref{fig:contmaps}e for the 357\,GHz emission.
The ellipses in the boxes at the bottom left and right corners show the
synthesized beam sizes of the 357\,GHz continuum emission image
($\theta_\mathrm{maj}\times\theta_\mathrm{min}\,=\,2\farcs 1\times 1\farcs 3$ 
at P.A.\,$=\,57^\circ$; Table \ref{tbl:obs}) 
and
the \NtwoH\ \Tex\ map
($4\farcs 6\times 3\farcs 7$ 
at P.A.\,$=\,-35^\circ$; Paper I), respectively.
See Appendix \ref{as:Tex} for details.
\textcolor{black}{(b)(right) Magnification of the central region of the plot shown in panel (a).
The cyan and blue contours, respectively, present the 3\sgm\ level ones
of the blue and red outflow lobes shown in Figure \ref{fig:lobemaps}.
}
\label{fig:Tex}}

\end{figure}

In \S\ref{ss:cont}, we used the mean excitation temperature
of the \NtwoH\ (1--0) emission over a circular region with a radius of 
\Reff\ $=$ \ReffValueUnit\ 
centered at the submm source [\meanTex\ $=\,22.6\pm 3.6$\,K].
The mean value was obtained from the \Tex\ map (Figure \ref{fig:Tex}) 
which was used to produce the column density 
map of \NtwoH\ shown in Figure 9b of Paper I.
\textcolor{black}{
In addition, we measured  the mean \Tex\ of
20.4$\pm$3.0\,K for the redshifted outflow lobes, including the southwestern island, 
and 17.4$\pm$5.6\,K for the blue one.
Here these values were calculated over the regions inside the 3\sgm\ level contours of 
the outflow lobes (see Figure \ref{fig:Tex}b).
}

\clearpage

\clearpage

\begin{deluxetable}{lllllccccrcl}
\tabletypesize{\scriptsize}
\tablecaption{Summary of the Interferometric Observations\label{tbl:obs}}
\tablewidth{0pt}
\tablehead{
\colhead{} & \colhead{} & \colhead{} & \colhead{} & \colhead{} &\colhead{} & \colhead{Spatial Frequency} &  \colhead{} & \multicolumn{2}{c}{Synthesized Beam} &  \colhead{} \\
\cline{9-10}
\colhead{Array} & \colhead{Emission} & \colhead{$f_\mathrm{cent}$} & \colhead{$\Delta f$\tablenotemark{a}} & \colhead{$\lambda$\tablenotemark{b}} & \colhead{FoV\tablenotemark{c}} & \colhead{Range} & \colhead{LAS\tablenotemark{d}} & \colhead{$\theta_\mathrm{maj}\times\theta_\mathrm{min}$} & \colhead{P.A.} & \colhead{Sensitivity} \\
\colhead{} & \colhead{} & \colhead{(GHz)} & \colhead{(MHz)} & \colhead{(mm)} & \colhead{(arcsec)} & \colhead{($k\lambda$)} & \colhead{(arcsec)} & \colhead{(arcsec)} & \colhead{(deg)} & \colhead{(mJy \pbeam)} \\
}
\startdata
CARMA & Continuum & 91.181 & 938 & 3.29 & 88  & 1.93 -- 101.7 &  107 & 3.89\,$\times$\,3.34 & $-84$ & 0.32  \\
SMA & Continuum & 267.755 &  3960 & 1.12 & 38 &  10.3 -- 102.7  &  20 & 2.63\,$\times$\,1.61 & 67 & 0.88 \\
SMA & Continuum & 279.755 &  3960 & 1.07 & 37 & 10.8 -- 107.3 & 19 &  2.57\,$\times$\,1.14 &  65 & 0.87 \\
SMA & Continuum & 344.820 &  3960 & 0.869 & 30 &  10.0 -- 131.8 & 21 &  2.20\,$\times$\,1.43 & 54 &1.6  &  \\
SMA & Continuum & 356.820 &   3960 & 0.840 & 29 & 7.7 -- 136.4 & 27 &  2.05\,$\times$\,1.34 & 57 & 1.7 &  \\
SMA & $^{12}$CO(3--2) & 345.796 & 88.4 & 0.867 & 30  & 7.7 -- 132.3 &  27 & 2.13\,$\times$\,1.39 & 55 & 122\tablenotemark{e}  &\\
\enddata
\tablenotetext{a}{\,Total bandwidth per polarization.}
\tablenotetext{b}{\,Wavelength.}
\tablenotetext{c}{\,Field-of-view.}
\tablenotetext{d}{\,Largest detectable angular size scale.}
\tablenotetext{e}{\,Sensitivity per velocity channel width of 0.4 \kms.}
\end{deluxetable}

\begin{deluxetable}{rclcccccc}
\tabletypesize{\scriptsize}
\tablecaption{Summary of the Interferometric Photometry of the Continuum Emission\label{tbl:phot}}
\tablewidth{0pt}
\tablehead{
\colhead{\lw{Frequency}} & \colhead{\lw{$\Theta_\mathrm{maj}\times\Theta_\mathrm{min}$\tablenotemark{a}}}  & 
\colhead{\lw{P. A.\tablenotemark{b}}} &
\colhead{\lw{$R_\mathrm{eff}$\tablenotemark{c}}} & \colhead{\lw{$I_\nu^\mathrm{peak}$\tablenotemark{d}}} & 
\multicolumn{2}{c}{$S_\nu$\tablenotemark{e}} & 
\colhead{\lw{$M_\mathrm{csm}$\tablenotemark{h}}} \\
\cline{6-7}
\colhead{} & \colhead{} &\colhead{}  & \colhead{} & \colhead{} & \colhead{2D Gaussian\tablenotemark{f}} & \colhead{3$\sigma$\tablenotemark{g}} & \colhead{} \\
\colhead{(MHz)} &\colhead{(arcsec$\times$arcsec)} & \colhead{(degree)} & \colhead{(AU)} 
& \colhead{(mJy \pbeam)} & \colhead{(mJy)} & \colhead{(mJy)} & \colhead{(\Msun)}  \\
}
\startdata
 89964\tablenotemark{i}  &   
                 6.62\,$\times$\,4.80  & \textcolor{black}{140$\pm$15} & 680$\pm$150 &  1.06$\pm$0.52  & 1.3$\pm$0.23 & \textcolor{black}{0.71$\pm$0.02} & 0.052$\pm$0.09 \\
 91181  &  7.56\,$\times$\,2.75  & \textcolor{black}{95$\pm$15} & 550$\pm$450 &  1.62$\pm$1.0 & 3.5$\pm$1.9 & \textcolor{black}{1.23$\pm$0.28\tablenotemark{j}}  & 0.13$\pm$0.07 \\
267755 &  2.60\,$\times$\,1.61  & 71$\pm$3 & 250$\pm$50 &   25.2$\pm$1.6 & 25.2$\pm$0.87 & \textcolor{black}{22$\pm$9} & 0.019$\pm$0.006 \\
279755 &  2.46\,$\times$\,1.71  & 61$\pm$4 & 250$\pm$50  &  29.6$\pm$1.7  & 33.2$\pm$2.0 & \textcolor{black}{30$\pm$12}  & 0.022$\pm$0.001 \\
344820 &  2.87\,$\times$\,1.43  & 54$\pm$3 & 240$\pm$110  &  28.6$\pm$2.5 & 36.6$\pm$3.4 & \textcolor{black}{34$\pm$9}  & 0.012$\pm$0.001 \\
356820 &  2.59\,$\times$\,1.80  & 81$\pm$12 & 260$\pm$90  &   19.7$\pm$2.7 & 33.4$\pm$5.2 & \textcolor{black}{24$\pm$6}  &  0.010$\pm$0.002 \\ 
\enddata
\tablecomments{See \S\ref{ss:cont} for details.}
\tablenotetext{a}{\,Beam-deconvolved source size obtained by task \texttt{JMFIT} in AIPS package with an assumption that 
the intensity distribution of the source is approximated by a 2D elliptical Gaussian
whose FWHMs along the major and minor axes are
$\Theta_\mathrm{maj}$ and $\Theta_\mathrm{min}$, respectively.}
\tablenotetext{b}{\,Position angle.}
\tablenotetext{c}{\,Effective source radius in AU.
$R_\mathrm{eff}$ is calculated from 
$\frac{\pi}{4\ln 2}(\Theta_\mathrm{maj}\times\Theta_\mathrm{min})\,d^2\,=\,\pi R_\mathrm{eff}^2$
where $d$ is the distance to the source in pc.
The uncertainty is estimated from the difference caused by the
elliptical and circular approximations.}
\tablenotetext{d}{\,Peak intensity obtained from the 
2D elliptical Gaussian fitting. 
The error in the fitting is calculated from the RMS noise level of each image (Table \ref{tbl:obs}).}
\tablenotetext{e}{\,Total flux density.}
\tablenotetext{f}{\,$S_\nu$ obtained from the 2D elliptical Gaussian fitting 
\textcolor{black}{by deconvolving the synthesized beam (see Table \ref{tbl:obs}.)}.}
\tablenotetext{g}{\,$S_\nu$ obtained by \textcolor{black}{integrating the emission over the area enclosed by the 3$\sigma$ level contour}.}
\tablenotetext{h}{\,``Circumstellar" mass estimated from the $S_\nu$ values in the column 6.}
\tablenotetext{i}{\,Continuum image was taken from Paper I. 
All the values are obtained from the re-analysis with the same method
as that of the other band data.}
\tablenotetext{j}{\,\textcolor{black}{Except for the weak emission elongated to the east
(see Figure \ref{fig:contmaps}a).}}
\end{deluxetable}

\begin{deluxetable}{lccrccccccccccccc}
\tabletypesize{\scriptsize}
\rotate
\tablecaption{Physical Properties of the Outflow Lobes\label{tbl:outflow}}
\tablewidth{0pt}
\tablehead{
\colhead{} & \colhead{\lw{$l/\sin i$\tablenotemark{a}}} & \colhead{\lw{P.A.\tablenotemark{b}}} & \colhead{\lw{$\theta$\tablenotemark{c}}} & \colhead{\lw{$v_\mathrm{flow}/\cos i$\tablenotemark{d}}} & \colhead{\lw{$\tau_\mathrm{dyn}\frac{\cos i}{\sin i}$\tablenotemark{e}}} & 
\multicolumn{2}{c}{$M_\mathrm{lobe}$\tablenotemark{f}} & & \multicolumn{2}{c}{$\dot{M}_\mathrm{outflow}\frac{\sin i}{\cos i} $\tablenotemark{g}} & & \multicolumn{2}{c}{$F_\mathrm{outflow}   \frac{\sin i}{\cos^2 i}$\tablenotemark{h}}
& & \multicolumn{2}{c}{$L_\mathrm{outflow}   \frac{1}{\cos^2 i}$\tablenotemark{i}} \\
\cline{7-8}\cline{10-11}\cline{13-14}\cline{16-17}
\colhead{\lw{Lobe}} & \colhead{} & \colhead{} & \colhead{} & \colhead{} & \colhead{} & \colhead{7.3\,K} & \colhead{22.6\,K} & & \colhead{7.3\,K} & \colhead{22.6\,K} & & \colhead{7.3\,K} & \colhead{22.6\,K} & & \colhead{7.3\,K} & \colhead{22.6\,K} \\
\colhead{} & \colhead{(AU)} & \colhead{(deg)} & \colhead{(deg)} & \colhead{(\kms)} & \colhead{(yrs)} & \multicolumn{2}{c}{($10^{-4}$\Msun)} & & \multicolumn{2}{c}{($10^{-8}$\Msun \,yr$^{-1}$)} & & \multicolumn{2}{c}{($10^{-7}$\Msun \,yr$^{-1}$\,yr$^{-1}$)} & & \multicolumn{2}{c}{($10^{-2}$ \Lsun)}  \\
}
\startdata
Blue: Northeast & 1600 & $\sim -55$   & $\sim 65$ & 5 & 2000 & 1.2 & 0.14 & & 5.5 & 0.71 & & 2.9 & 0.37 & & 1$\times 10^{-2}$ & 2$\times 10^{-3}$\\ 
Blue: Southwest & 1300 & $\sim 125$ & $\sim 85$ &  5 & 1700  & 2.2 & 0.28 & & 13 & 0.17 & & 6.8 & 0.88 & & 3$\times 10^{-2}$ & 4$\times 10^{-3}$\\
\textcolor{black}{Red:  IV} & 3600 & $\sim 130$  & $\sim 110$ &  11 & 2100  & 44 & 5.6 & & 200 & 26 & & 230 & 30 & & 2 & 0.3 \\
Red:  HV & 1000 & $\sim 130$ & $\sim 70$  & 13 & 500 & 17 & 0.22 & & 34 & 4.4 & & 46 & 5.9 & & 0.5 & 6$\times 10^{-2}$ \\
\enddata
\tablenotetext{a}{\,Outflow lobe length calculated from the measured lobe length 
seen in Figures \ref{fig:lobemaps} and \ref{fig:redlobemaps}. Hence we corrected for
the unknown inclination angle of the outflow axis ($i$) by assuming $i \,=\,45^\circ$.}
\tablenotetext{b}{\,Outflow lobe position angle measured by eye. The uncertainty is typically 10\degr.}
\tablenotetext{c}{\,Outflow lobe opening angle measured by eye. The uncertainty is typically 30\degr.}
\tablenotetext{d}{\,Outflow velocity given by 
$v_\mathrm{flow} = |v_\mathrm{t}-v_\mathrm{sys}|/\cos i$.}
\tablenotetext{e}{\,Dynamical time scale given by $\tau_\mathrm{dyn} = l/v_\mathrm{flow}$.}
\tablenotetext{f}{\,Outflow lobe mass obtained by the \CO\ integrated intensity 
over the velocity range defined in \S\ref{sss:outflow}.
\textcolor{black}{
The left and right column values correspond to the masses in the cases of 
\Tex\ $=$ 7.3\,K and 22.6\,K, respectively.
}}
\tablenotetext{g}{\,Outflow mass loss rate estimated by
$\dot{M}_\mathrm{outflow} = M_\mathrm{lobe}/\tau_\mathrm{dyn}$.}
\tablenotetext{h}{\,Outflow momentum rate estimated by
$F_\mathrm{outflow} = M_\mathrm{lobe}v_\mathrm{flow}/\tau_\mathrm{dyn}$.}
\tablenotetext{i}{\,Outflow mechanical luminosity estimated by
$L_\mathrm{outflow} = \frac{1}{2}M_\mathrm{lobe}v_\mathrm{flow}^2$.}
\end{deluxetable}

\begin{deluxetable}{cccc}
\tabletypesize{\scriptsize}
\tablecaption{Summary of the Continuum Photometry with Cameras\label{tbl:apphot}}
\tablewidth{0pt}
\tablehead{
\colhead{Wavelength} & \colhead{\lw{Camera}} & \colhead{Aperture} & \colhead{$S_\nu$} \\ 
\colhead{(\micron)} & & \colhead{(arcsec)} & \colhead{(mJy)} 
}
\startdata
\hline 
350 &  SHARC-II & 8.4 & $\lesssim 1830$\tablenotemark{a} \\
70 &  MIPS & 18.0 & $\lesssim 940$\tablenotemark{a} \\
24 &  MIPS &  6.0 & 184.0\tablenotemark{c} \\
22.09 & WISE4 & 12.0 & $238.5 \pm 5.4$\tablenotemark{d} \\
11.56 & WISE3 &  6.5 & $14.693 \pm 0.028$\tablenotemark{d} \\
8.0 & IRAC & 1.98 & 11.7\tablenotemark{c} \\
5.8 & IRAC & 1.88 & 6.4\tablenotemark{c} \\
4.60 & WISE2 & 6.4 & $3.080 \pm 0.0059$\tablenotemark{d} \\
4.5 & IRAC & 1.72 & 3.1\tablenotemark{c} \\
3.6 & IRAC & 1.66 & 0.72\tablenotemark{c} \\
3.35 & WISE1 & 6.1 & $(2.691 \pm 0.079)\times 10^{-2}$\tablenotemark{d} \\\hline 
\enddata
\tablecomments{See \S\ref{ss:IRcont_results} and Figure \ref{fig:SpitzerMaps} for the Spitzer results.}
\tablenotetext{a}{\,Considered to be an upper limit because the aperture is significantly larger than the source size.}
\tablenotetext{b}{\,Taken from Paper I.}
\tablenotetext{c}{\,Photometric uncertainties in Spitzer images are typically 10\%.}
\tablenotetext{d}{\,Photometry at the WISE bands are taken from those for \texttt{WISE~J205129.83+601838} 
in the WISE All-Sky Release Source Catalog.}
\end{deluxetable}

\begin{deluxetable}{rlcc}
\tabletypesize{\scriptsize}
\tablecaption{Summary of the SED model fits\label{tbl:sedfitter}}
\tablewidth{0pt}
\tablehead{
\colhead{\lw{Property}} & \colhead{\lw{Symbol}}  & \colhead{\lw{Unit}}  & \colhead{\lw{Results\tablenotemark{a}}} \\ 
}
\startdata
\hline 
Stellar radius & $R_\ast$ & \Rsun\ & 3.7 \\
Stellar surface temperature &  $T_\ast$ & K &  3400 \\
Disk mass [dust] & $M_\mathrm{disk}^\mathrm{dust}$ & \Msun\ & $7.6\times 10^{-3}$ \\
Disk outer radius\tablenotemark{c} & $R^\mathrm{disk}_\mathrm{max}$ &  AU   & 50  \\
Disk flaring power & $\beta_\mathrm{disk}$ & $\cdot\cdot\cdot$ &  1.1\\
Disk surface density power & $p$ & $\cdot\cdot\cdot$ & $-1.7$ \\
Disk scaleheight & $h$ & AU  & 3.7 \\
Envelope density [dust] &  $\rho_0^\mathrm{env}$ & g\,cm$^{-3}$ &  $1.5\times 10^{-24}$ \\
Cavity density [dust] & $\rho_0^\mathrm{cav}$ & g\,cm$^{-3}$ &  $5.1\times 10^{-23}$  \\
Cavity opening angle & $\theta_0$ &  deg &  20 \\
Cavity power & $c$ & $\cdot\cdot\cdot$  & 1.4 \\
Inclination angle & $i$ & deg  & \textcolor{black}{65} \\
\enddata
\tablecomments{Results from SED fitting using YSO models in \citet{Robitaille17}.
See \S\ref{ss:SED} for details.}
\tablenotetext{a}{\,See Figure \ref{fig:sedfitter}. The best-fit model is \texttt{g5QAQXBF\_07} in the model set of \texttt{spubhmi}.}
\tablenotetext{b}{\,Corresponds to envelope inner radius.}
\end{deluxetable}

\end{document}